\documentclass[twocolumn]{aastex631}

\usepackage{natbib}
\usepackage[version=4]{mhchem}
\usepackage{enumitem}

\definecolor{background}{RGB}{252, 229, 241}
\definecolor{frame}{RGB}{225, 34, 138}

\newlist{gitemize}{itemize}{4}
\setlist[gitemize,1]{
  leftmargin=\dimexpr1.5cm+\labelsep\relax,
  label={\smash{\raisebox{-0.25\height}{\includegraphics[width=0.6cm]{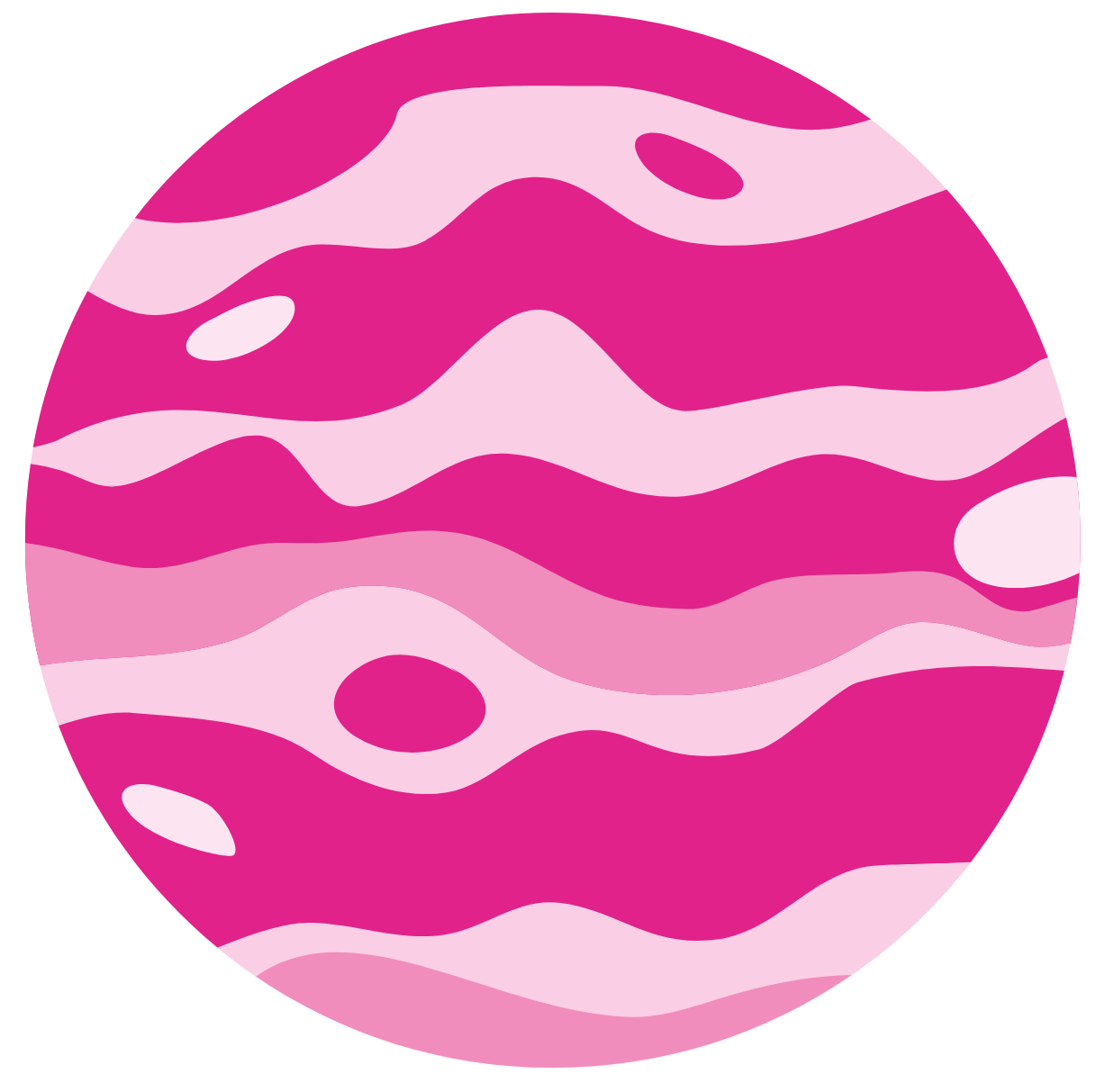}}}}
}

\newcommand{\RNum}[1]{\uppercase\expandafter{\romannumeral #1\relax}}

\DeclareSymbolFont{UPM}{U}{eur}{m}{n}
\DeclareMathSymbol{\umu}{0}{UPM}{"16}
\let\oldumu=\umu
\renewcommand\umu{\ifmmode\oldumu\else\math{\oldumu}\fi}
\newcommand\micro{\umu}

\let\microns \jmicron

\shorttitle{Introducing the KEN}
\shortauthors{Latouf et al.}

\begin{document}
\title{Bayesian Analysis for Remote Biosignature Identification on exoEarths (BARBIE) \RNum{3}: \\Introducing the KEN}

\author[0000-0001-8079-1882]{Natasha Latouf}
\altaffiliation{NSF Graduate Research Fellow, 2415 Eisenhower Ave, Alexandria, VA 22314}
\affiliation{Department of Physics and Astronomy, George Mason University, 4400 University Drive MS 3F3, Fairfax, VA, 22030, USA}
\affiliation{NASA Goddard Space Flight Center, 8800 Greenbelt Road, Greenbelt, MD 20771, USA}
\affiliation{Sellers Exoplanents Environment Collaboration, 8800 Greenbelt Road, Greenbelt, MD 20771, USA}

\author[0000-0002-9338-8600]{Michael D. Himes}
\affiliation{NASA Postdoctoral Program Fellow, NASA Goddard Space Flight Center, 8800 Greenbelt Road, Greenbelt, MD 20771, USA}
\affiliation{Morgan State University, 1700 E Cold Spring Lane, Baltimore, MD 21251, USA}
\affiliation{NASA Goddard Space Flight Center, 8800 Greenbelt Road, Greenbelt, MD 20771, USA}

\author[0000-0002-8119-3355]{Avi M. Mandell}
\affiliation{NASA Goddard Space Flight Center, 8800 Greenbelt Road, Greenbelt, MD 20771, USA}
\affiliation{Sellers Exoplanents Environment Collaboration, 8800 Greenbelt Road, Greenbelt, MD 20771, USA}

\author[0000-0001-7912-6519]{Michael Dane Moore}
\affiliation{NASA Goddard Space Flight Center, Greenbelt, MD, USA.}
\affiliation{Business Integra, Inc., Bethesda, MD, USA.}
\affiliation{Sellers Exoplanets Environment Collaboration, 8800 Greenbelt Road, Greenbelt, MD 20771, USA}

\author[0000-0002-5060-1993]{Vincent Kofman}
\affiliation{NASA Goddard Space Flight Center, 8800 Greenbelt Road, Greenbelt, MD 20771, USA}
\affiliation{Sellers Exoplanents Environment Collaboration, 8800 Greenbelt Road, Greenbelt, MD 20771, USA}
\affiliation{Integrated Space Science and Technology Institute, Department of Physics, American University, Washington DC}

\author[0000-0002-2662-5776]{Geronimo L. Villanueva}
\affiliation{NASA Goddard Space Flight Center, 8800 Greenbelt Road, Greenbelt, MD 20771, USA}
\affiliation{Sellers Exoplanents Environment Collaboration, 8800 Greenbelt Road, Greenbelt, MD 20771, USA}

\author{Chris Stark}
\affiliation{NASA Goddard Space Flight Center, 8800 Greenbelt Road, Greenbelt, MD 20771, USA}
\affiliation{Sellers Exoplanents Environment Collaboration, 8800 Greenbelt Road, Greenbelt, MD 20771, USA}

\correspondingauthor{Natasha Latouf}
\email{nlatouf@gmu.edu, natasha.m.latouf@nasa.gov}

\begin{abstract}
%7,434,720
We deploy a newly-generated set of geometric albedo spectral grids to examine the detectability of methane (\ce{CH4}) in the reflected-light spectrum of an Earth-like exoplanet at visible and near-infrared wavelengths with a future exoplanet imaging mission. By quantifying the detectability as a function of signal-to-noise ratio (SNR) and molecular abundance, we can constrain the best methods of detection with the high-contrast space-based coronagraphy slated for the next generation telescopes such as the Habitable Worlds Observatory (HWO). We used 25 bandpasses between 0.8 and 1.5 {\microns}. The abundances range from a modern-Earth level to an Archean-Earth level, driven by abundances found in available literature. We constrain the optimal 20\%, 30\%, and 40\% bandpasses based on the effective SNR of the data, and investigate the impact of spectral confusion between \ce{CH4} and \ce{H2O} on the detectability of each one. We find that a modern-Earth level of \ce{CH4} is not detectable, while an Archean Earth level of \ce{CH4} would be detectable at all SNRs and bandpass widths. Crucially, we find that \ce{CH4} detectability is inversely correlated with \ce{H2O} abundance, with required SNR increasing as \ce{H2O} abundance increases, while \ce{H2O} detectability depends on \ce{CH4} abundance and selected observational wavelength, implying that any science requirements for the characterization of Earth-like planet atmospheres in the VIS/NIR should consider the abundances of both species in tandem.

\end{abstract}


\section{Introduction}
\label{sec:intro} 

The Habitable Worlds Observatory (HWO), recommended by the \citet{decadal}, is slated to launch in the 2040s with the foundational goal of constraining the properties of Earth-like exoplanets using high-contrast imaging. This type of imaging is one of the first steps on the path to determining planet habitability for potentially Earth-like worlds and providing context for the development of our own Earth through time. The \citet{decadal} identified a primary science driver to detect and characterize 25 exoEarth candidates (EECs) in the habitable zones (HZ) of nearby stars. Characterization of an exoplanet's atmosphere can provide vital information on the formation, history, and physical composition of the planet; for potentially habitable planets, we can also search their atmospheres for biosignatures that can hint at the likelihood of clement conditions and biological activity \citep{schwieterman18}. A holistic understanding of robust biological indicators and necessary signal-to-noise ratio (SNR) for detection is crucial to establishing an efficient observing procedure and driving instrument development for HWO.

%However, an efficient observation strategy to optimize biosignature detection across the range of physically plausible atmospheric properties through wavelength remains unconstrained, although there is progress being made on specific atmospheric constituents in specific wavelength regimes. 
%Further, quantifying the detector requirements to detect biosignatures is still in a phase of infancy. 
%Key fundamental questions include: 
% \begin{gitemize}
%     \item \textit{How does biosignature detectability vary as a function of spectral band position, planetary archetype, and stellar host? }
%     \item \textit{What is the most efficient observing procedure to characterize multiple biosignatures?}
%     \item \textit{What are the requirements on coronagraph performance in order to maximize exoplanet yields with the desired biosignature detectability?}
% \end{gitemize}

With the advances in high-contrast imaging instrumentation planned for HWO \citep{luvoir}, the ability to detect flux from a habitable Earth-twin is becoming a realistic possibility, which can unlock new molecular detections. Previous works, such as \citet{feng18,damiano23}, and \citet{young24}, have investigated biosignature detectability at varying wavelengths, such as \ce{O3} in the UV, and found that detectability is drastically effected by the chosen observational wavelength.
%varying the wavelength of observation to other wavelength regimes drastically effects the detectability.
There are many molecules that peak in their absorption outside of the optical region, and would thus require a more thorough investigation of required resolving power and abundance for detection.

Methane (\ce{CH4}) shows absorption in the visible and NIR wavelength regimes, with a range of spectral features varying in optical depth. Other molecules such as \ce{O2} and \ce{O3} can have abiotic production mechanisms and thus may not necessarily indicate a biosignature \cite{schindler2000, domagalgoldman14}. Even if from biotic sources, many eras of early Earth had extremely low levels of \ce{O2}, such as the Proterozoic, leading to a very unlikely or impossible chance of detection \cite{planavsky14,latouf24}. However, in low-\ce{O2} eras of Earth, it is possible that the atmosphere was instead \ce{CH4}-rich, such as in the Archean era, and thus \ce{CH4} would be readily detectable \cite{arney16, wogan2020}. Further, \ce{CH4} has a very short lifetime in Earth's atmosphere when only produced by abiogenic sources due to destruction by oxygen, and thus a detection of a high \ce{CH4} abundance would strongly indicate biogenic sources as well as abiogenic \citep{krissansen18, wogan2020}. Enabling a detection of \ce{CH4} can increase the likelihood of confirming an Earth-twin discovery. 

\begin{figure*}[ht!]
\centering
\includegraphics[scale=0.5]{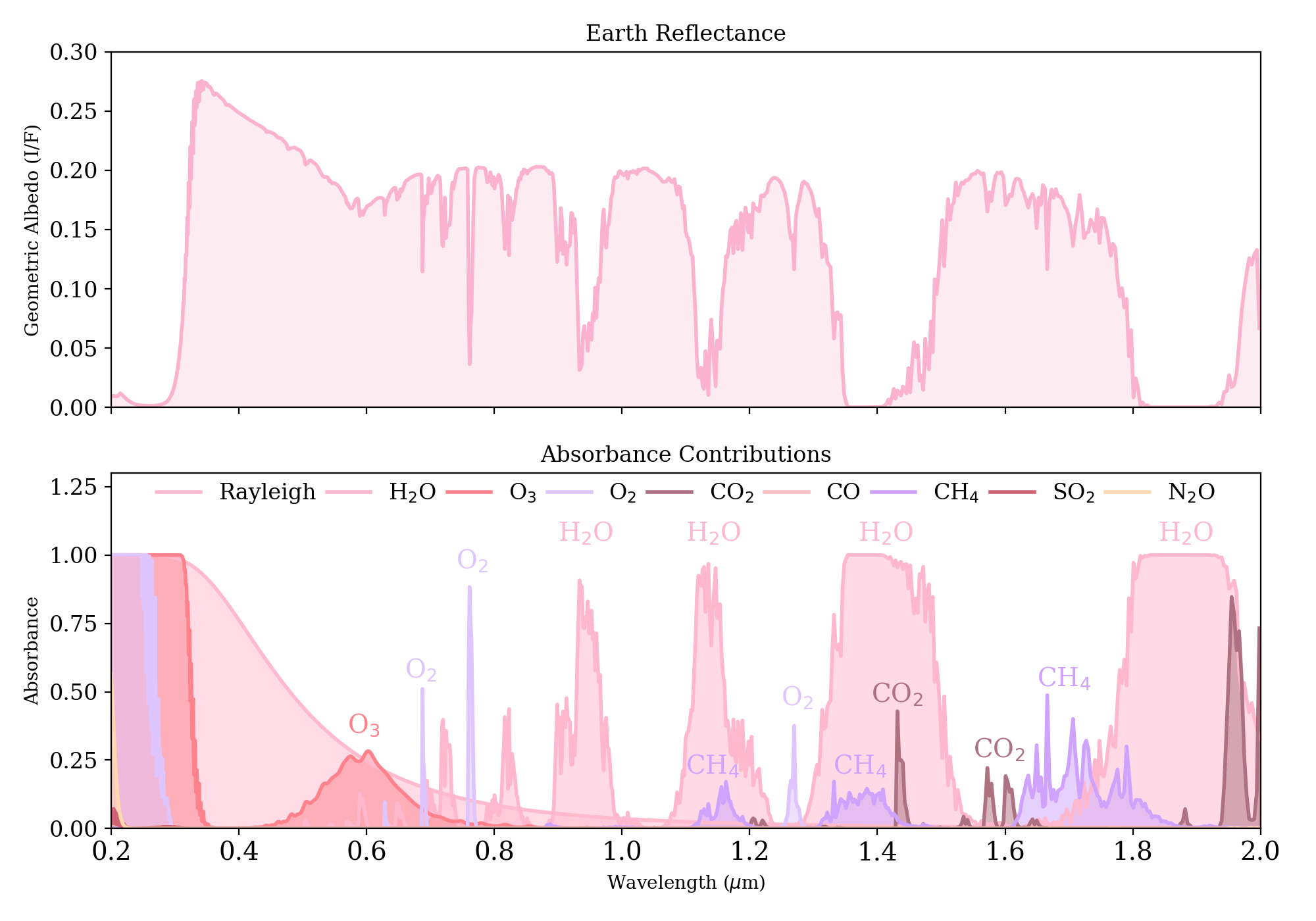}
\caption{In the top panel, we present a representative spectrum from our work where molecular contributions are visible, with the y-axis representing the geometric albedo. In the bottom panel, we present the absorbance contributions of every molecule available in the KEN grids, as well as the Rayleigh scattering also included: \ce{H2O}, \ce{O2}, \ce{O3}, \ce{CH4}, \ce{CO}, \ce{CO2}, \ce{SO2}, and \ce{N2O}, with the y-axis representing absorption. The x-axes are both the full wavelength range of the KEN grids (0.2--2{\microns}). Note that \ce{SO2} and \ce{N2O} absorb at 0.2 {\microns} very minimally, without features in other locations in this wavelength range.}
\label{fig:fullspec_opt}
\end{figure*}

In this work, we use the spectral grid-based Bayesian inference method first described in \citet[hereafter S23][]{susemiehl23}, and used in \citet[hereafter BARBIE1][]{latouf23} and \citet[hereafter BARBIE2][]{latouf24}. In this method, 1.4 million geometric albedo spectra at discrete parameter values are generated, and forward models for Bayesian retrievals are then interpolated from the grid. However, the S23 grid has a limited wavelength range, operating primarily at optical wavelengths (0.5--1 {\microns}), and consists of six varied parameters: surface pressure ($\mathrm{P_{0}}$), surface albedo ($\mathrm{A_{s}}$), and gravity (g), \ce{H2O}, \ce{O2}, and \ce{O3}. However, the limited set of atmospheric constituents bounded the applicability of the S23 grid to oxygen-rich atmospheric compositions at visible wavelengths (0.4 - 1.0), and necessitated new grids to explore new wavelength ranges and atmospheric compositions. Figure~\ref{fig:fullspec_opt} portrays a representative spectrum for an Earth-twin in the top panel, and every molecule and their absorbance across the full spectral range of the new grids (0.2--2 {\microns}) in the bottom panel. By focusing only on the visible, many molecular species are not considered, including key biosignatures such as \ce{CH4} and \ce{CO2}. By building new grids with larger wavelength coverage and more parameters, we can further investigate the optimization of exoplanet characterization observations and build a more robust observing strategy for biosignatures.

% \begin{figure*}[h!]
% \centering
% \includegraphics[scale=0.5]{optical_withfullspec_allmol.png}
% \caption{Four reflection spectra: in shades of pink and purple, varying \ce{H2O} abundances limited only to the optical range (0.5--1 {\microns}) prescribed by the S23 grid. In dotted black, the full wavelength range of the KEN grids (0.2--2 {\microns}) with all available parameters: \ce{H2O}, \ce{O2}, \ce{O3}, \ce{CH4}, \ce{CO}, \ce{CO2}, \ce{SO2}, and \ce{N2O}. All values in the dotted black are set to modern-Earth abundances.}
% \label{fig:fullspec_opt}
% \end{figure*}

This project is a direct continuation from BARBIE1 and BARBIE2, and in this paper we extend the same methodology to validate our new grids and study the impact of longer wavelengths (i.e. into the NIR) with \ce{CH4}. In $[\S]$ \ref{sec:method} we present the methodology of our grid-building, validation, and simulations, also providing a brief summary of BARBIE1 and BARBIE2. In $[\S]$ \ref{sec:results} we present the results of our simulations for modern and varying abundances of \ce{CH4}, as well as an analysis of how \ce{CH4} and \ce{H2O} detectability affect one another. In $[\S]$ \ref{sec:disc} we discuss the presented results and analyze the impact for future observations of varying Earth-twin epochs as a function of wavelength and further bandpass widths. In $[\S]$ \ref{sec:conc} we present our conclusions and ideas for future work.

\section{Methodology}
\label{sec:method}

\subsection{A New Grid-Building Scheme}
\label{sec:grids}  

Although spectral retrieval studies have been used extensively to explore the detectability of atmospheric compositions for the direct imaging of exoplanets for multiple telescopes both current and future \citep[e.g.,][]{lupu16, nayak17, smith20, damiano22,young24}, these retrievals are extremely computationally expensive. Most Bayesian retrievals use real-time radiative transfer calculations, combined with exploring large parameter spaces, mission capabilities, and constant model improvement that increases computation time to a rate that is not easily usable. Different methods to accelerate retrievals have been explored through efficient radiative transfer schemes or machine learning \cite[e.g.,][]{robinson22, zingales18, neila18, cobb19, fisher20, HimesEtal2022}. 

In order to facilitate the efficient production of new optimized spectral grids for our grid-based retrievals, we developed a Python package called Gridder, which is a generalized grid-building scheme based on the methodology of S23 \citep{himes24}. Gridder produces arbitrary spectral grids using the Planetary Spectrum Generator, a publicly available radiative transfer model for creating planetary spectra \citep[PSG,][]{PSG,PSGbook}. PSG can be used to calculate spectra over an ultra-broad wavelength range (50 nm to 100 mm) and includes planetary atmospheres, surfaces, and bulk properties such as aerosols, atomic, continuum, and molecular scattering/radiative processes implemented layer-by-layer. The Gridder parameter structure is highly customizable, allowing any input parameter to PSG to be used as a parameter in the optimized grid, including parameterized thermal profiles (e.g., isothermal, adiabatic, \citet{line2013}) and atmospheric chemistry (e.g., constant-with-altitude, thermochemical equilibrium). In addition to the error metric of S23, Gridder offers other common metrics, such as the maximum absolute percent error and mean squared error, for greater control over the resulting grid's accuracy. Gridder features various optimizations (e.g., parallelization, checkpoints) to ensure efficient production of optimized spectral grids in a consistent, reproducible manner. Using Gridder, we have built a new set of spectral grids, each with six parameters to decrease grid-building computational time, but over a far wider wavelength range from the UV to NIR (0.2--2.0 \microns), and we have added the additional molecular atmospheric constituents \ce{CH4}, \ce{CO2}, \ce{CO}, \ce{SO2}, and \ce{N2O}.

\begin{figure*}[ht!]
\centering
\includegraphics[width=\textwidth]{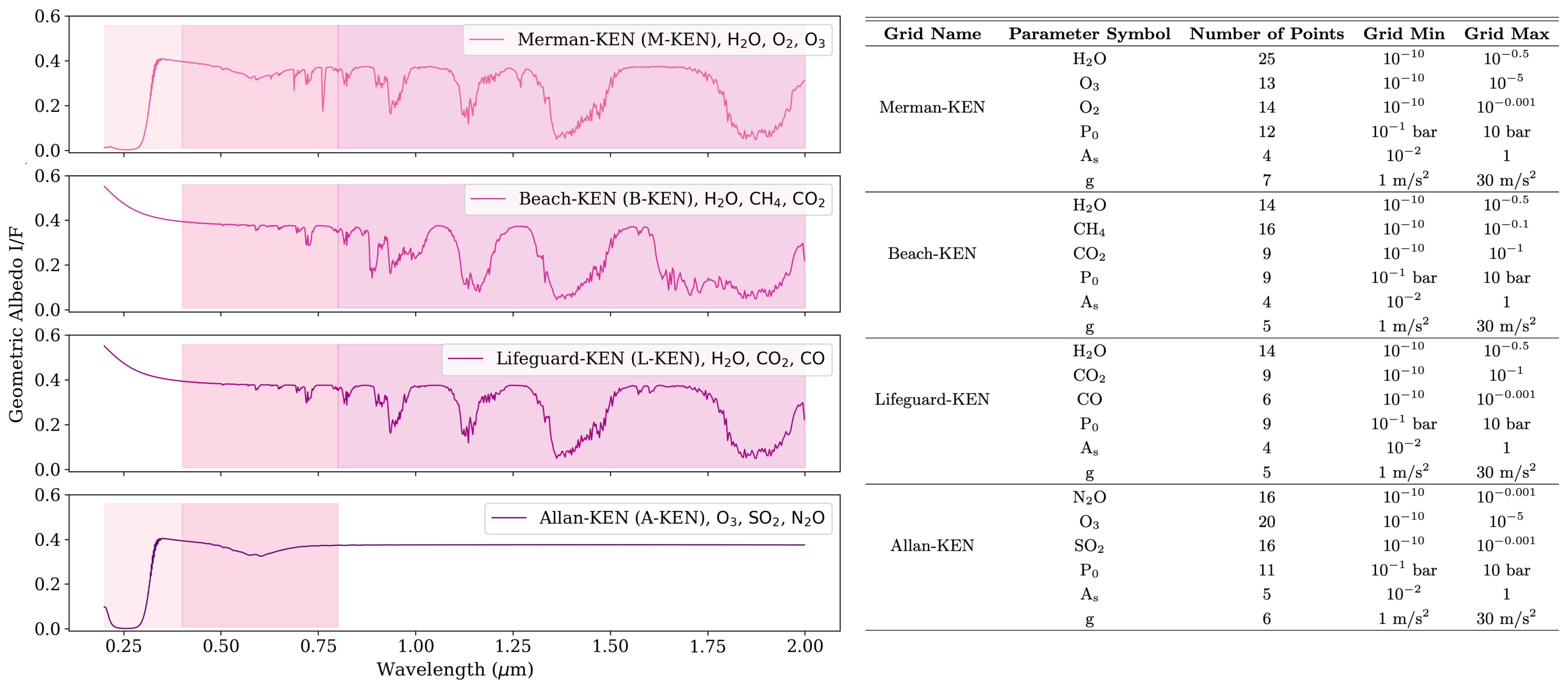}
\caption{Herein we present a modern, 50\% cloudy, Earth-like spectrum for each grid over the full wavelength range, to illustrate the absence and presence of each molecule per grid. Each legend shows the name of the grid along with the present molecular constituents, with geometric albedo on the y-axis and wavelength on the x-axis. The highlighted regions portray the UV, Visible, and NIR wavelength regions, and they are present in the grids that are most useful in said region. The table on the right contains the grid structure for all KEN grids (grid name, spacing, and number of points) for the clear grid versions. For the combined clear and cloudy grids, the overall size is doubled. It also includes the minimum and maximum values for each grid parameter.}
\label{fig:allgrids}
\end{figure*}

Our newly built grid set, hereafter the KEN grids (no acronym; they’re just Ken), allow for a direct comparison between grids since they cover the same wavelength range and three base parameters of $\mathrm{P_{0}}$, $\mathrm{A_{s}}$, and g with $\mathrm{R_p}$ fixed to 1 $R_\Earth$. However, they reduce computational time by taking advantage of the different spectral regions covered by absorption features of different molecules. Any molecules that have overlapping spectral features (for the range of atmospheric pressures examined) are housed within the same grid, while any that are orthogonal to each other do not need to be simultaneously retrieved within one grid, such as \ce{CH4} and \ce{O3}. The background gas in each grid is set to \ce{N2} = 1 - $P_{1}$ - $P_{2}$ - $P_{3}$ where $P_{1}$ -- $P_{3}$ are the molecules per grid. In this way, we can also investigate the effects of false positive or negative molecular detection due to overlapping features. We developed four different spectral grids at a resolving power of 500 (R = 500) which can then be binned to lower resolving powers, with the ability to vary the cloud fraction $\mathrm{C_{f}}$ at any value between 0\% cloudy to 100\% cloudy. We describe the molecular parameters in each named grid in the KEN grid set, along with a modern Earth-like spectrum per grid and the highlighted wavelength regimes for use per grid, in Figure~\ref{fig:allgrids}.

The KEN grids will allow us to investigate varying atmospheric compositions beyond \ce{O2}-dominated atmospheres, such as \ce{CH4} or \ce{CO2} dominated, to develop observational procedures for multiple planetary archetypes through wavelength.

% Due to the higher error possible at grid edges, leading to inaccurate retrieval results, we ensure the lowest values in our parameter space are far below those that are of interest to investigate, although the results of BARBIE1 and BARBIE2 inform the abundances at which \ce{H2O}, \ce{O3}, and \ce{O2} are detectable.

\subsection{BARBIE Methodology}
\label{sec:barbie}
    
We follow an identical methodological approach to that of BARBIE1 and BARBIE2 for the \ce{CH4} case study analysis. Herein we present a brief summary of the main steps in our analysis. For more detailed information, please see BARBIE1 and BARBIE2. 

\begin{enumerate}
    \item We set a modern-Earth twin as our fiducial data spectrum following \citet{feng18}, with isotropic volume mixing ratios (VMRs)  \ce{H2O}$=3\times10^{-3}$, \ce{O3}$=7\times10^{-7}$, \ce{O2}=$0.21$, constant temperature profile at 250 K, $\mathrm{A_{s}}$ of 0.3, $\mathrm{P_{0}}$ of 1 bar, and a planetary radius fixed at $\mathrm{R_p}$ = 1 $\mathrm{R_\Earth}$. We bin our grid from the native resolving power of 500 to 140 and 70 for our simulations, and split the spectrum in 25 evenly spaced bandpasses across 20\%, 30\%, and 40\% widths. R = 140 is generally accepted in the optical regime, while R = 70 is generally accepted in the NIR \citep{luvoir}. These ranges mimic the simultaneous bandpasses that may be achieved with high-performance coronagraphs in the future, or dual coronagraphs that can be used in combination \citep{ruane15, por20, roser22}.
    \item We run a series of Bayesian nested sampling retrievals using PSGnest\footnote{https://psg.gsfc.nasa.gov/apps/psgnest.php} housed in PSG. PSGnest is a Bayesian spectral retrieval methodology adapted from the original Fortran version of the Multinest retrieval algorithm \citep{multinest} and designed for application to exoplanetary observations (i.e. incorporating grid multi-dimensional interpolation, grid retrievals, the fitting methods, as well as the Multinest retrievals).
    \item We receive the highest-likelihood values, the average values from the posterior distributions, uncertainties, and the log evidence (logZ) \citep{PSGbook} as outputs. We then calculate the log-Bayes factor \citep[$\mathrm{lnB}$;][]{benneke13}; this compares retrievals with and without a given molecule to estimate the likelihood of the molecule’s presence in the exoplanet’s atmosphere. For our purposes, $\mathrm{lnB}<2.5$ is unconstrained (no detection), $2.5\le\mathrm{lnB}<5.0$ is a weak detection, and $\mathrm{lnB}{\ge}5.0$ is a strong detection \citep[see Table 2 of][]{benneke13}. We also calculate the median, upper, and lower limit values of the 68\% credible region \citep{harrington22}. However $\mathrm{lnB}$ is a better estimation of detection rather than the 68\% credible region since it is directly investigating the molecular presence vs absence and calculating which iteration yields the best fit, while the 68\% credible region, by definition, includes the true value 68\% of the time. $\mathrm{lnB}$ is not calculated for non-gaseous components, since those factors cannot be absent.
\end{enumerate}

\begin{table}
    \centering
        \centering
        \begin{tabular}{cc}
            \hline
            \hline
            \textbf{\ce{CH4} (VMR)} & \textbf{Assumed Earth Epoch (Age)} \\
            \hline
            1.65$\times10^{-6}$ & Hadean (3.9 Gyr)\\
            % 1$\times10^{-5}$ & Filler\\
            % 7$\times10^{-5}$ & Filler\\
            % 1$\times10^{-4}$ & Filler\\
            1.65$\times10^{-3}$ & Archean (3.5 Gyr)  \\
            7.07$\times10^{-3}$ & Archean (2.4 Gyr) \\
            1.65$\times10^{-3}$ & Proterozoic (2.0 Gyr)\\
            4.15$\times10^{-4}$ & Phanerozoic (0.8 Gyr)\\
            1.65$\times10^{-6}$$^{\dagger}$ & Phanerozoic (0.3 Gyr)$^{\dagger}$ \\
            \hline
        \end{tabular}
        \caption{The above values were used in our simulations for \ce{CH4}, moving from Earth-like values into different epochs of Earth's history based on \cite{kasting2005} and \citet{kaltnegger07}, and interpolating to provide more points between large gaps of abundance. \\$^{\dagger}$ Modern Earth-like value.}
        \label{tab:valsch4}
    \end{table}
    
This process is then repeated for varying abundances. For this work, we vary \ce{CH4} according to Table~\ref{tab:valsch4}, following \citet{kasting2005} and \citet{kaltnegger07}, while also adding values to bridge the large gap between abundance values (i.e., between Hadean and Archean values). The level of \ce{CH4} present in any specific habitable exoplanet atmosphere is uncertain, since \ce{CH4} production is significantly driven by biological activity; the same is true for any other biologically-produced species. Since we are not running fully physically-consistent chemistry models, we therefore use modern-Earth values as the default for other species as well as to isolate the effect of a single molecule, as in prior BARBIE works. We also investigate the effects of \ce{H2O} on \ce{CH4} detectability by testing a range of combined \ce{H2O} and \ce{CH4} values, shown in Table~\ref{tab:h2och4combo}. All retrievals were performed between 0.8 and 1.5 {\microns} with 20\%, 30\%, and 40\% bandpass widths, and R=140 and R=70. {We carefully select a narrower wavelength range in order to investigate the detectability of \ce{CH4} in the optical or at the critical 0.9 {\microns} \ce{H2O} feature; this is currently poised as the first step in observations following the decision tree technique in \citet{young24}.

% We then add another layer of complexity by adding coronagraph detector QE curves to our noise estimation, while also varying the intrinsic SNR point-by-point within a given bandpass. This gives us a more accurate view of the instrument influence on molecular detectability, especially in for \ce{CH4} and \ce{H2O} which are in a critical inflection point between the optical and NIR detector QE.

\begin{figure*}
\centering
\includegraphics[width=\textwidth]{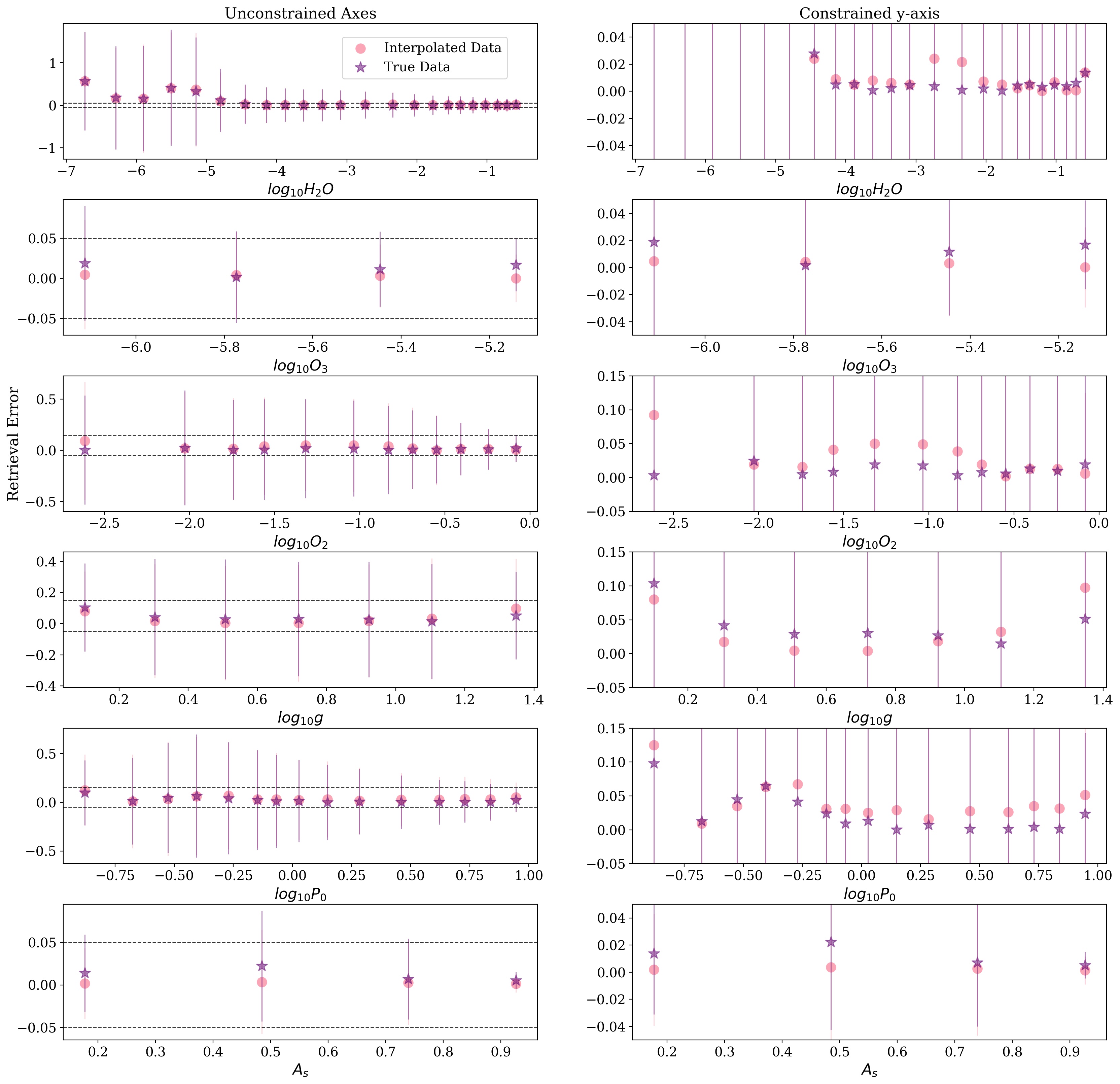}
\caption{Interpolation error across a 6D retrieval across parameter space. The retrieval error is defined as the difference between the true and retrieved values. The pink dots represent retrievals performed on a spectrum interpolated from the KEN grid, and the purple stars represent retrievals performed on a PSG-calculated spectrum. The error bars on these points are drawn as the 68\% credible regions. The left column shows the full y-axes, and the right column zooms in to the portion on the y-axes contained within the horizontal dashed black lines that cover the areas of lowest error.}
\label{fig:interp_nick}
\end{figure*}

\section{Results}
\label{sec:results}

\subsection{Grid Validation}
\label{sec:gridval}

In S23 Figure 8, they investigate the interpolation error in the developed grid to ensure that the grid is not introducing large error past any known interpolation error. We have replicated this in Figure~\ref{fig:interp_nick} using the same technique as in S23 using our M-KEN grid, as that has the same parameters as in the S23 grid. To summarize, we ran 6D retrievals for each value of each parameter in the test grid. The test grid in use consists of the midpoint values of the M-KEN grid as interpolation error is typically maximized near the midpoints, and thus to test the error we want to investigate the area of highest error. For each parameter set, a ``true'' spectrum was created (i.e., pulling the data spectrum directly from the midpoints grid) at an SNR of 20. The retrievals were then repeated, but the data spectrum was created by interpolating from the main M-KEN grid (i.e., the way that all retrievals will work when using KEN grids with PSGnest). We use the offset between the values retrieved for the ``true" and interpolated spectra to assess the impact of interpolation error on these retrievals.

Although the M-KEN grid covers a wider wavelength range than the S23 grid, Figure~\ref{fig:interp_nick} covers the wavelength range of 0.4--1 {\microns} in order to directly replicate the range in S23. The error from retrieving on interpolated data is shown in pink dots, while the error from retrieving on``true'' data is shown in purple stars. We also calculate the 68\% credible region and portray those as error bars. The left column shows the full extent of the offset for every point, while the right column zooms in to the regions shown in the black dotted lines. We select these areas based on where there is a high concentration of overlapping points. Any differences between the results are due to interpolation error. 

We find that the retrieval error is generally very similar for both the ``true'' and interpolated data retrievals, with the derived retrieval error for both always within $1\sigma$ of the input value. The outlier errors, as in S23, are at small $\mathrm{P_{0}}$, and at large \ce{O3}, although even those error ranges are no more than 15\%. The average error hovers at below 5\%, while the error between the interpolated and true data is never larger than 2\% (with the singular exception of 10\% at low \ce{O3}). In fact, our error spread is much smaller than S23, validating that the new optimization procedures of Gridder yield higher fidelity results with a much faster overall runtime. This confirms that the Gridder selection of grid points has been successful, with interpolation error not serving as a significant inhibitor for performing retrievals in this region. 

%% I think the analysis I performed fits here since it is related to the above but over a wider wavelength range and higher resolution, but feel free to move it as you see fit

To further isolate the impact of the retrieval error induced by interpolation within the grid apart from the Bayesian inference routine, we performed 4800 retrievals using the UltraNest package \citep{Buchner2016statcompTestNestedSampling, Buchner2019paspCollabNestedSampling, Buchner2021jossUltraNest} on PSG-simulated spectra for values near the midpoints across the B-KEN grid's parameter space.  We use the grid's native $R=500$ spectral resolution and the full 0.2--2 {\microns} wavelength range and assume a SNR of 15 for each wavelength channel.  We find that the mean absolute error (MAE) between the maximum likelihood and the known true value is typically ${\sim}$0.02 for $A_s$ and ${\sim}$0.2 for all other parameters when there is sufficient information content to constrain the parameter.  This finding is generally consistent with Figure~\ref{fig:interp_nick}.  When \ce{H2O}, \ce{CH4}, and \ce{CO2} can simultaneously be constrained, the MAE for the extrema of the 3${\sigma}$ region of all parameters is nearly equal to the MAE of the maximum likelihood, indicating that the 1D marginalized posteriors are offset from the known truth.  This bias is generally negligible except in the case of $g$, where the mean absolute percentage error can be $>60$\% at low values of $\mathrm{log}_{10}g$ and ${\sim}20$\% for Earth-like gravities.  These results represent worst-case scenarios; under typical grid usage conditions ($R=140$, reduced wavelength coverage, at least one parameter not at a grid midpoints), these errors will be reduced.

We present another analysis of grid efficacy in Figure~\ref{fig:archeancorner}, looking specifically at the deep \ce{CH4} feature centered at 1.1 {\microns}. This presents modern-Earth values of all parameters except \ce{CH4}, which is at an Archean value - increasing the \ce{CH4} value is necessary to investigate grid efficacy, as we expect it to be well constrained. This is using the B-KEN grid, which is also the grid used for all the science cases explored in this work. The B-KEN grid is the only KEN grid that contains \ce{CH4}, thus it is our only option to investigate \ce{CH4} detectability. We find that there are no sharp cut-offs in our 1D histograms except for $g$, which can especially be seen in the 2D marginalized posterior for $g$ and $\mathrm{P_{0}}$. However, $g$ around the Earth value is the primary use case for these grids, which is well retrieved. \ce{CO2} and $g$ also have significant posterior density at 1$\times10^{-10}$ VMR and 1 m/s$^{2}$ respectively, indicating that the 68\% credible region is underestimated. However, any molecule at such low abundances does effect spectral change, and \ce{CO2} is unconstrained in this wavelength region where there is no \ce{CO2}. We also see known and expected degeneracies with $\mathrm{A_{s}}$ and $\mathrm{C_{f}}$, with every parameter, especially the molecular parameters, heavily impacted, but every parameter is also retrieved within the 68\% credible region.

% However, assuming any \ce{CO2}< 1$\times10^{-10}$ VMR is equivalent to 1$\times10^{-10}$ VMR, which is appropriate since there will be effectively no spectral change at that small abundance, broadening the 68\% credible region and ensuring the truth is no longer at the edge of the credible region.
\begin{figure*}
\centering
\includegraphics[scale=0.25]{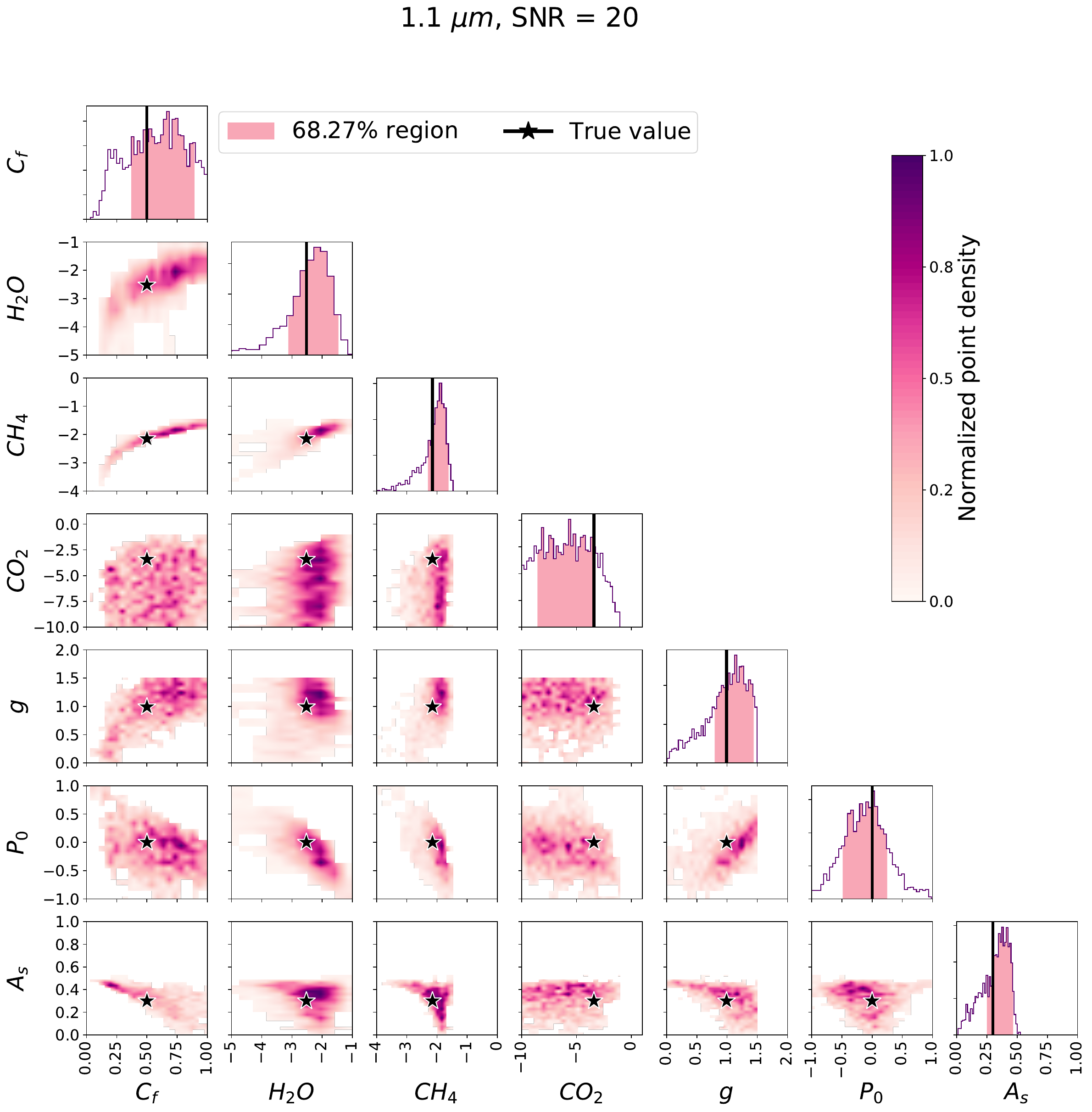}
\caption{Corner plot for Archean \ce{CH4} abundance. The 68\% credible regions are shown as pink shading in the 1D marginalized posterior distributions along the diagonal of the corner plot, and the true values are represented by black lines in the diagonals of the corner plot, and black stars within the 2D plots.}
\label{fig:archeancorner}
\end{figure*}

For further analysis on grid, and Gridder, validation, please see \citet{himes24} for a more thorough description.

\begin{figure*}[ht!]
\centering
\gridline{
          \fig{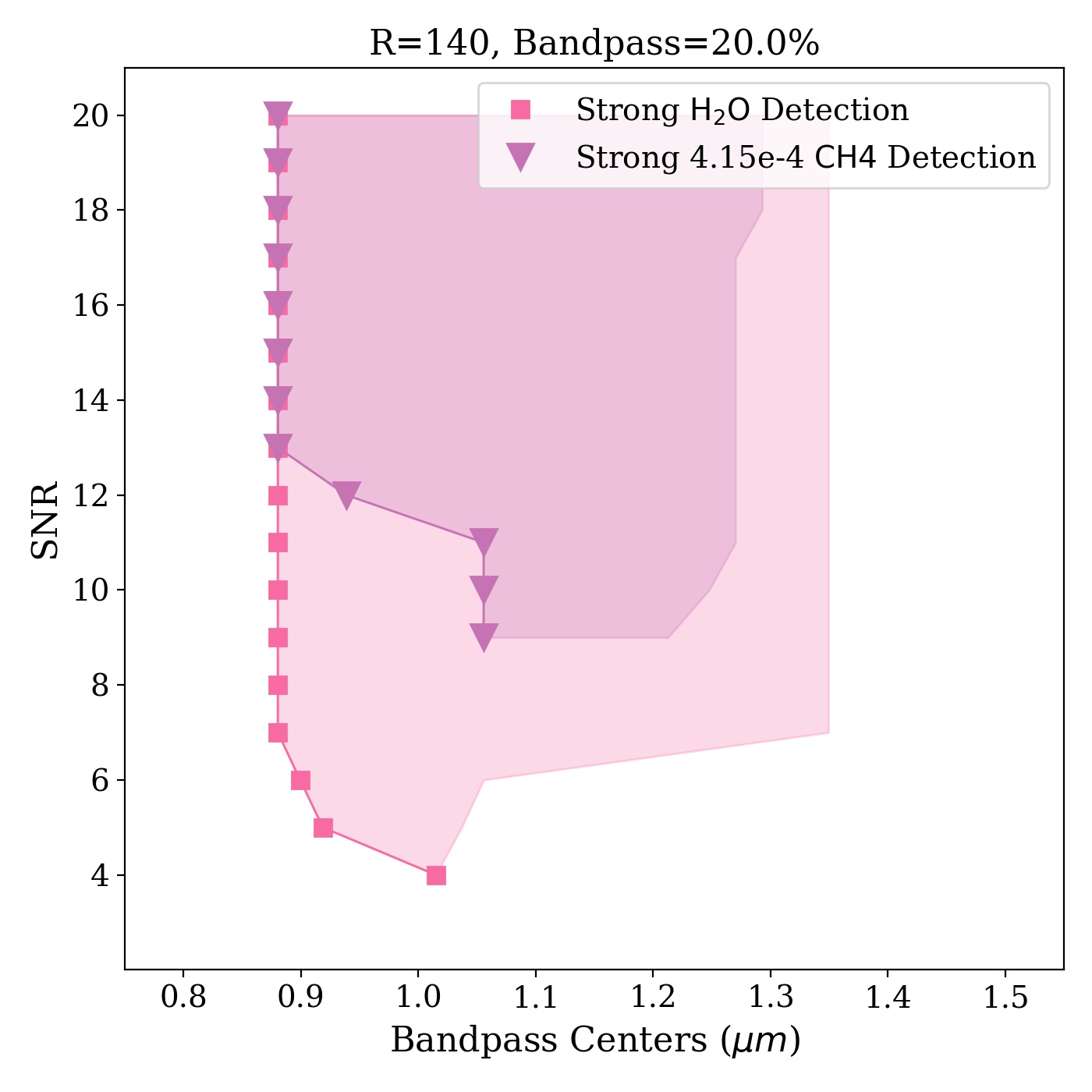}{0.335\textwidth}{(a) 4.15$\times10^{-4}$ \ce{CH4}}
          \fig{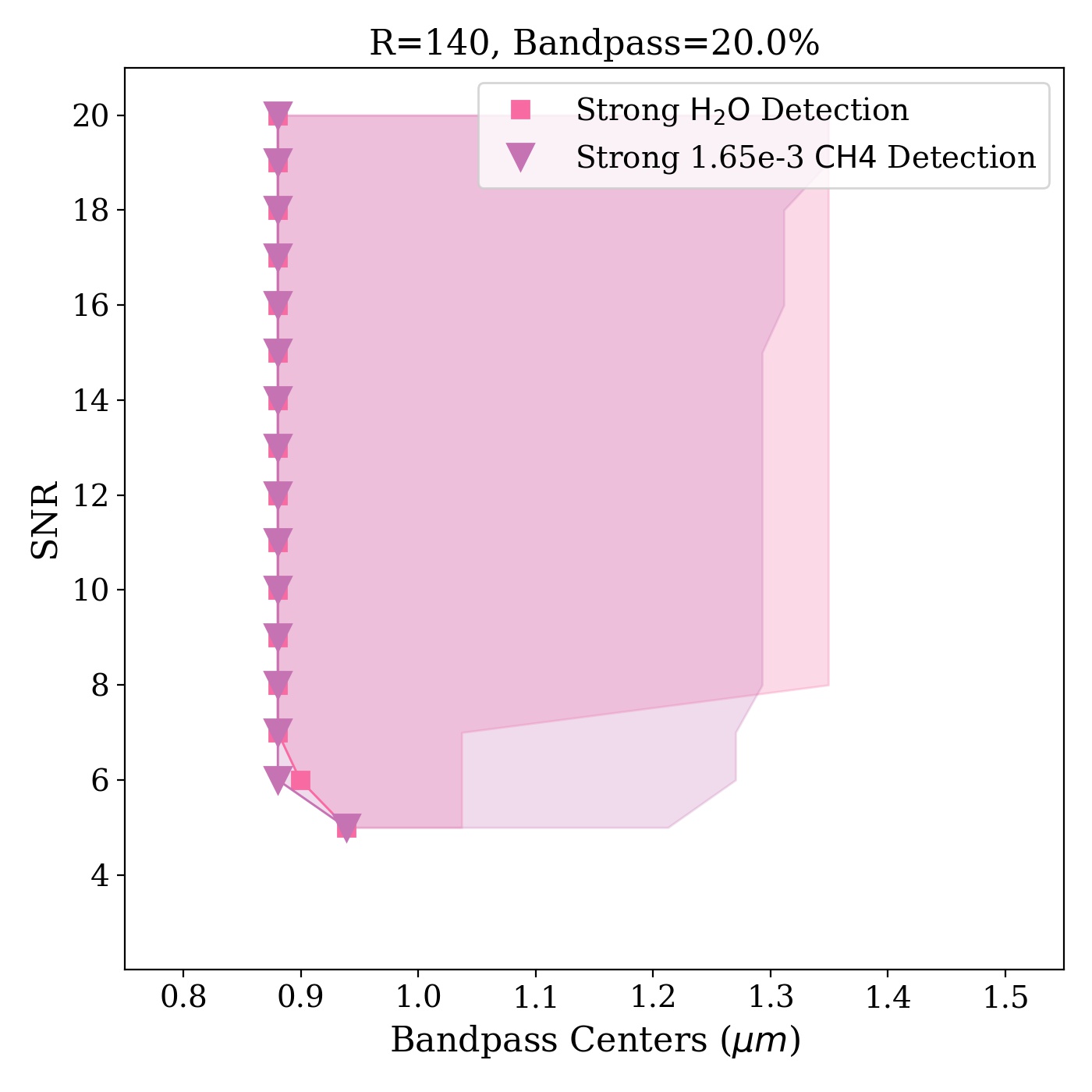}{0.335\textwidth}{(b) 1.65$\times10^{-3}$ \ce{CH4}}
          \fig{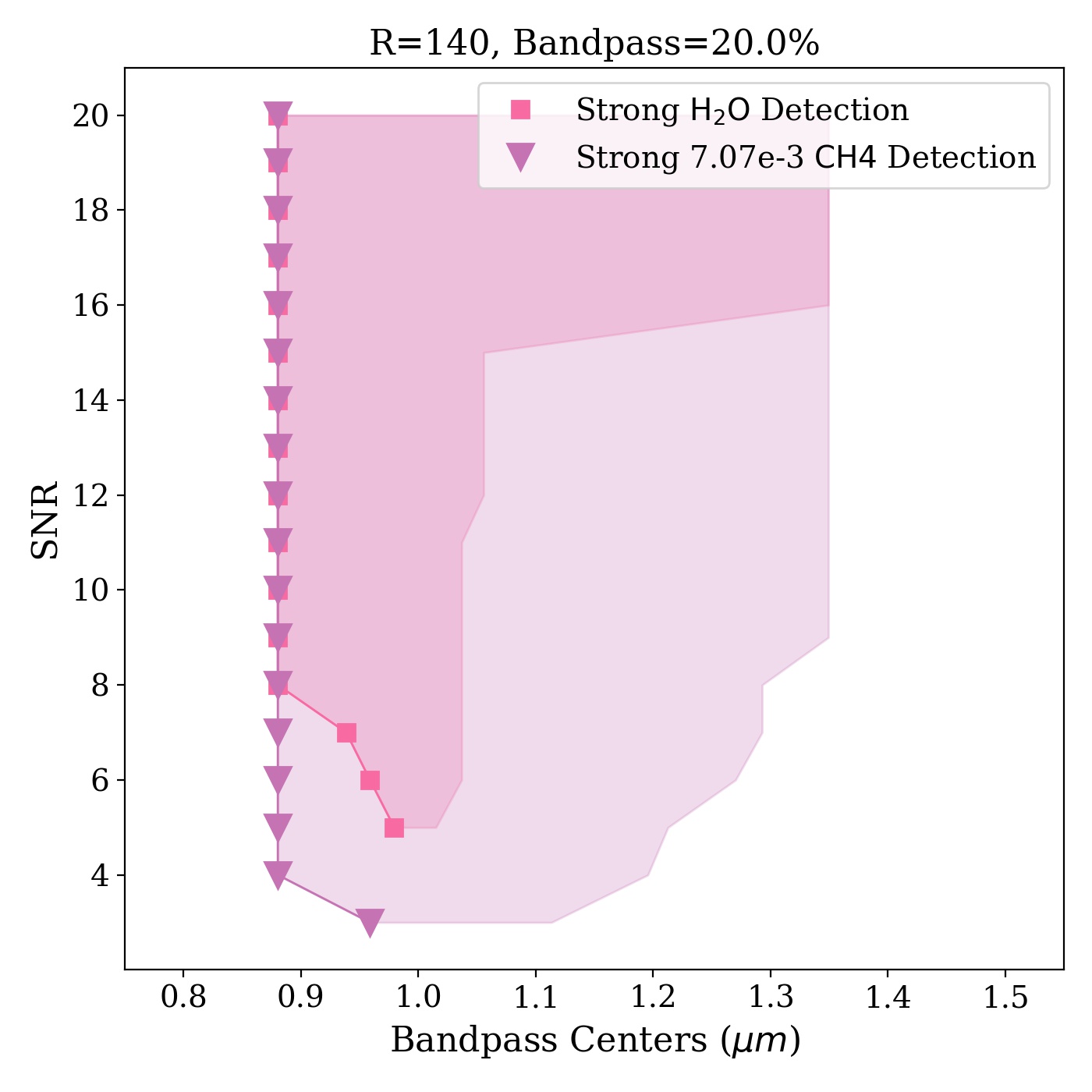}{0.335\textwidth}{(c) 7.07$\times10^{-3}$ \ce{CH4}}}
\caption{Summary of the bandpasses where \ce{CH4} and \ce{H2O} are detectable for 3$\times10^{-3}$ (modern) \ce{H2O} and varying \ce{CH4} based on Table~\ref{tab:valsch4}. The shortest bandpass center at which one can achieve a strong detection for \ce{H2O} or a strong detection for \ce{CH4}, are shown in pink squares and light purple triangles respectively. SNR is on the y-axis, and the bandpass centers are on the x-axis. We also show the range out to the longest wavelength at which the same detection is achieved as shaded regions.}
% \hspace{-5cm}
\label{fig:h2och4}
\end{figure*}

\subsection{Geologically Motivated \ce{CH4} Abundances}
\label{sec:ch4results}

We begin by presenting the detectability of \ce{CH4} as a function of SNR for the fiducial modern-Earth case, as examined by S23, BARBIE1, and BARBIE2, to maintain consistency (all \ce{CH4} data and calculated log-Bayes factors across abundance, SNR, and wavelength will be available to the community on Zenodo\footnote{DOI: 10.5281/zenodo.13760695}). All other parameters were left to modern Earth values, including a modern Earth value of \ce{H2O}. However,  a modern level of \ce{CH4} (1.65$\times10^{-6}$) is completely undetectable at at all SNRs and bandpass widths tested at a resolving power of 140 and 70. Many other works have previously found the same result, due to the low concentration and thus weak absorption features \citep{schwieterman18, damiano22, young24, samgj24}. Since a modern Earth \ce{CH4} was not possible to detect, we quickly proceeded to the varying abundances listed in Table~\ref{tab:valsch4} to explore the minimum value at which \ce{CH4} could be detected. 

For our abundance case study, we vary the abundance of \ce{CH4} above modern Earth values up to a cap of the Archean Earth value (7.07$\times10^{-3}$). We also vary the SNR for each abundance level to fully investigate the trade-off between higher SNR and molecular detectability. The \ce{CH4} abundances investigated are presented in Table~\ref{tab:valsch4}, along with intermediate values to fill in the large gap between the Hadean and Archean epochs. 
%We find that bandpass width has no impact on the detectability of \ce{CH4} in this region, due to our limited wavelength regime. From 0.8--1.5 {\microns}, a 20\% bandpass covers almost the full width of the spectral features, and thus increasing the bandpass width does not aid in the detection of molecules. Thus, we only present the 20\% bandpass width results. 

In Figures~\ref{fig:h2och4}a, \ref{fig:h2och4}b, and \ref{fig:h2och4}c, we present the strong detectability of the three highest \ce{CH4} abundances in our study, along with a modern \ce{H2O} abundance for context, in purple triangles and pink squares respectively. SNR is on the y-axis, with bandpass centers on the x-axis. We see that, as expected, as the abundance of \ce{CH4} increases, the required SNR for detection decreases, with SNRs of 9, 5, and 3 for \ce{CH4} abundances of 4.15$\times10^{-4}$, 1.65$\times10^{-3}$, and 7.07$\times10^{-3}$, respectively. Notably, at 4.15$\times10^{-4}$, the lowest SNR (9) is accessible at $\sim$ 1.05 {\microns}, where at 1.65$\times10^{-3}$ and 7.07$\times10^{-3}$ abundances, the lowest SNRs are accessible at $\sim$ 0.9 {\microns}. In order to achieve a detection of \ce{CH4} at 4.15$\times10^{-4}$ at 0.9 {\microns}, an SNR of 13 is necessary. 

Looking at the full range of abundance values in our study, we present a heat map showing detection strength as a function of SNR (y-axis) and \ce{CH4} abundance (x-axis) in Figure~\ref{fig:heatmapsnowater}a. The color bars represent the range of 0--5 log-Bayes factor, between unconstrained and strong detection. We see here that \ce{CH4} is not strongly detectable until 4.15$\times10^{-4}$ (Phanerozoic), at which point an SNR of 9 is required. Moving to higher abundances significantly drops the required SNR for strong detection, with 1.65$\times10^{-3}$ (Early Phanerozoic) requiring an SNR of 5 and 7.07$\times10^{-3}$ (Archean) detectable at all SNRs. At low abundances, such as a modern Earth value of 1.65$\times10^{-6}$, \ce{CH4} is not detectable at any SNR. Interestingly, at 1$\times10^{-4}$ a weak detection is possible at an SNR of 20, with the next value of 4.15$\times10^{-4}$ only requiring an SNR of 9 for strong detection as mentioned, and an SNR of 6 for weak detection. This results in a very drastic change in \ce{CH4} detection over a relatively small change in abundance of \ce{CH4}. 

However, returning to Figure~\ref{fig:h2och4}, we see a significant change in detectability for \ce{H2O} across \ce{CH4} abundance. A higher SNR is required to detect \ce{H2O} as \ce{CH4} increases. Also as the \ce{CH4} abundance increases, the range of strong \ce{H2O} detection as a function of wavelength decreases, with a sharp cut off at $\sim$ 1.1 {\microns}. For longer wavelengths than 1.1 {\microns}, it requires a high SNR ($\ge15$) to detect \ce{H2O} although there is a very large \ce{H2O} feature at 1.35 {\microns}. This indicates that the \ce{H2O} features are being masked by strong \ce{CH4} features, thus \ce{H2O} detectability is impacted.

\begin{figure*}[ht!]
\centering
\gridline{
          \fig{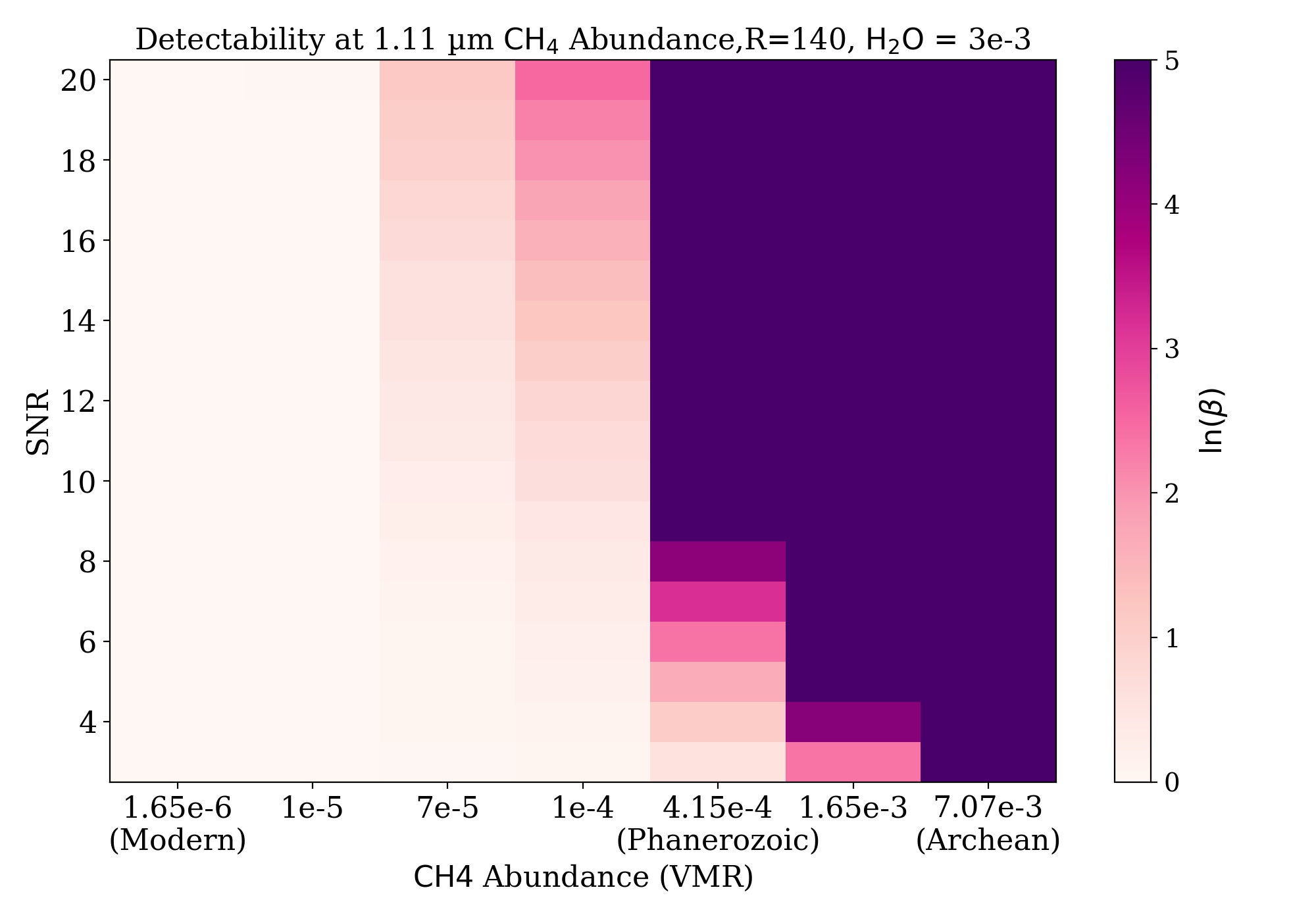}{0.5\textwidth}{(a) Modern \ce{H2O} Abundance}
          \fig{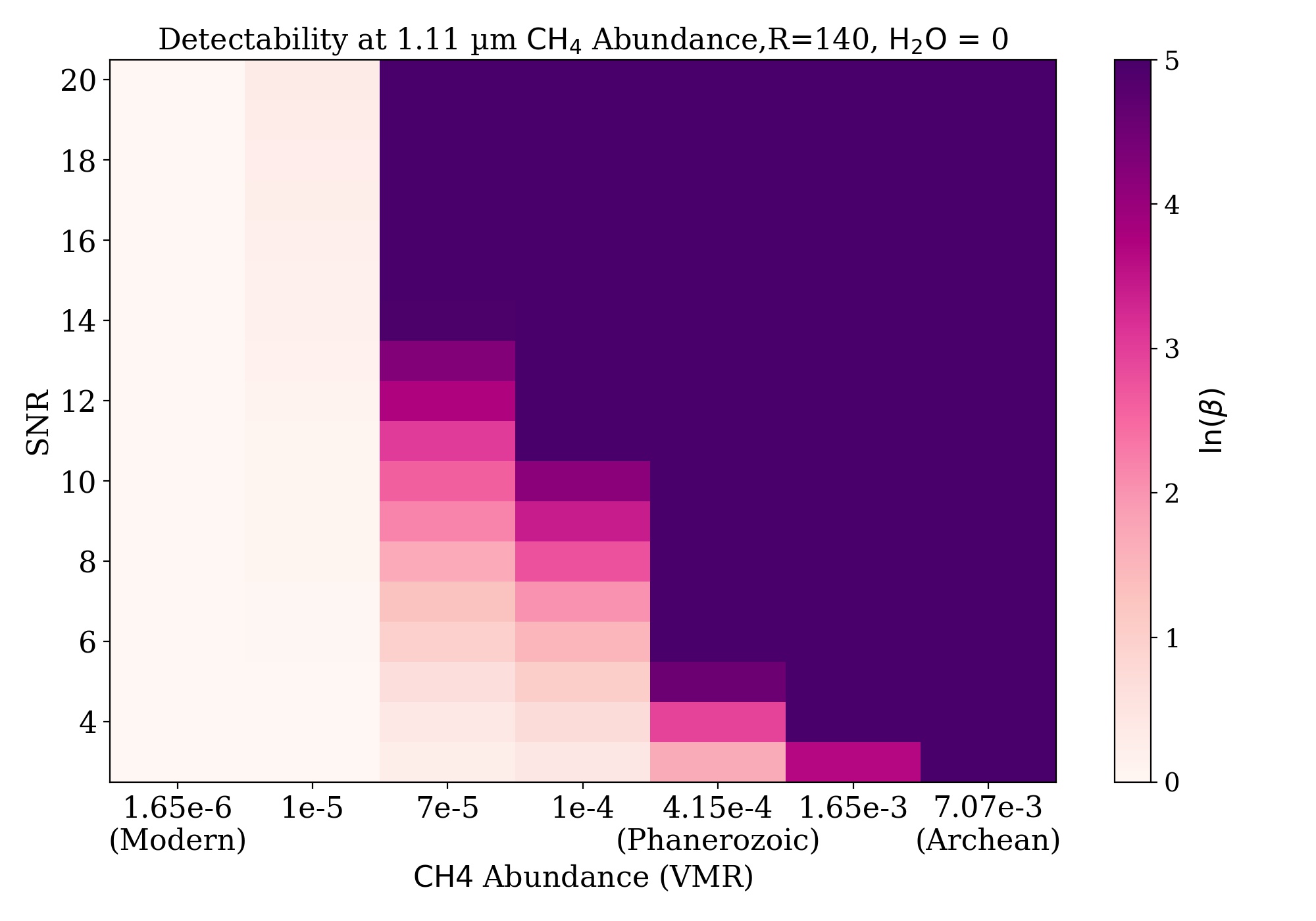}{0.5\textwidth}{(b) No \ce{H2O}}}
\caption{Heatmap plots illustrating detection strength as a function of SNR and varying \ce{CH4} abundance with two \ce{H2O} values. SNR is on the y-axis, \ce{CH4} abundance is on the x-axis, and the color bar shows the range of log-Bayes Factor ($\mathrm{lnB}$) from 0 to 5 to describe detection strength. $lnB<2.5$ are unconstrained, $2.5 \leq lnB < 5$ are weak, and $lnB>5$ are strong, as in prior figures.}
% \hspace{-5cm}
\label{fig:heatmapsnowater}
\end{figure*}

This is confirmed in Figure~\ref{fig:varych4modernh2o}, which presents the spectrum of modern \ce{H2O} and three increasing values of \ce{CH4} in three panels (Phanerozoic, Early, and Late Archean from top to bottom) as a function of wavelength and geometric albedo (x-axis and y-axis, respectively). As \ce{CH4} increases in abundance, the absorption feature aligns more closely with that of \ce{H2O}, particularly between $\sim$ 1.1 {\microns} and 1.2 {\microns}. This results in an inflection point, wherein \ce{H2O} is more difficult to detect due to the absorption feature being masked by the overpowering \ce{CH4}. This led us to investigate the opposite effect - does \ce{H2O} mask the detection of \ce{CH4} at lower \ce{CH4} abundance?

We reran all values of \ce{CH4} in Table~\ref{tab:valsch4}, however we set the abundance of \ce{H2O} to zero in our fiducial spectrum. We find that indeed, our detection of \ce{CH4} changes dramatically. In Figure~\ref{fig:heatmapsnowater}b, we again present a detectability heatmap across all abundances of \ce{CH4} as in Figure~\ref{fig:heatmapsnowater}a, although with no \ce{H2O} present. We see that \ce{CH4} is strongly detectable down to 7$\times10^{-5}$ VMR at an SNR of 14, and 4.15$\times10^{-4}$ VMR was detectable at an SNR of 6. Thus, with the absence of \ce{H2O}, our detectable \ce{CH4} abundance shifts down over an order of magnitude. We can see the effect of \ce{H2O} on \ce{CH4} detectability, and this can suggest that the abundance of \ce{CH4} cannot accurately be determined unless you also separately constrain \ce{H2O} abundance. However, we want to further understand the interactions between varying abundances of \ce{H2O} and \ce{CH4}. 

%certain abundances of \ce{CH4} can be used to indirectly infer the presence or absence of \ce{H2O} from the \ce{CH4} detection strength

\begin{figure*}
\centering
\includegraphics[scale=0.43]{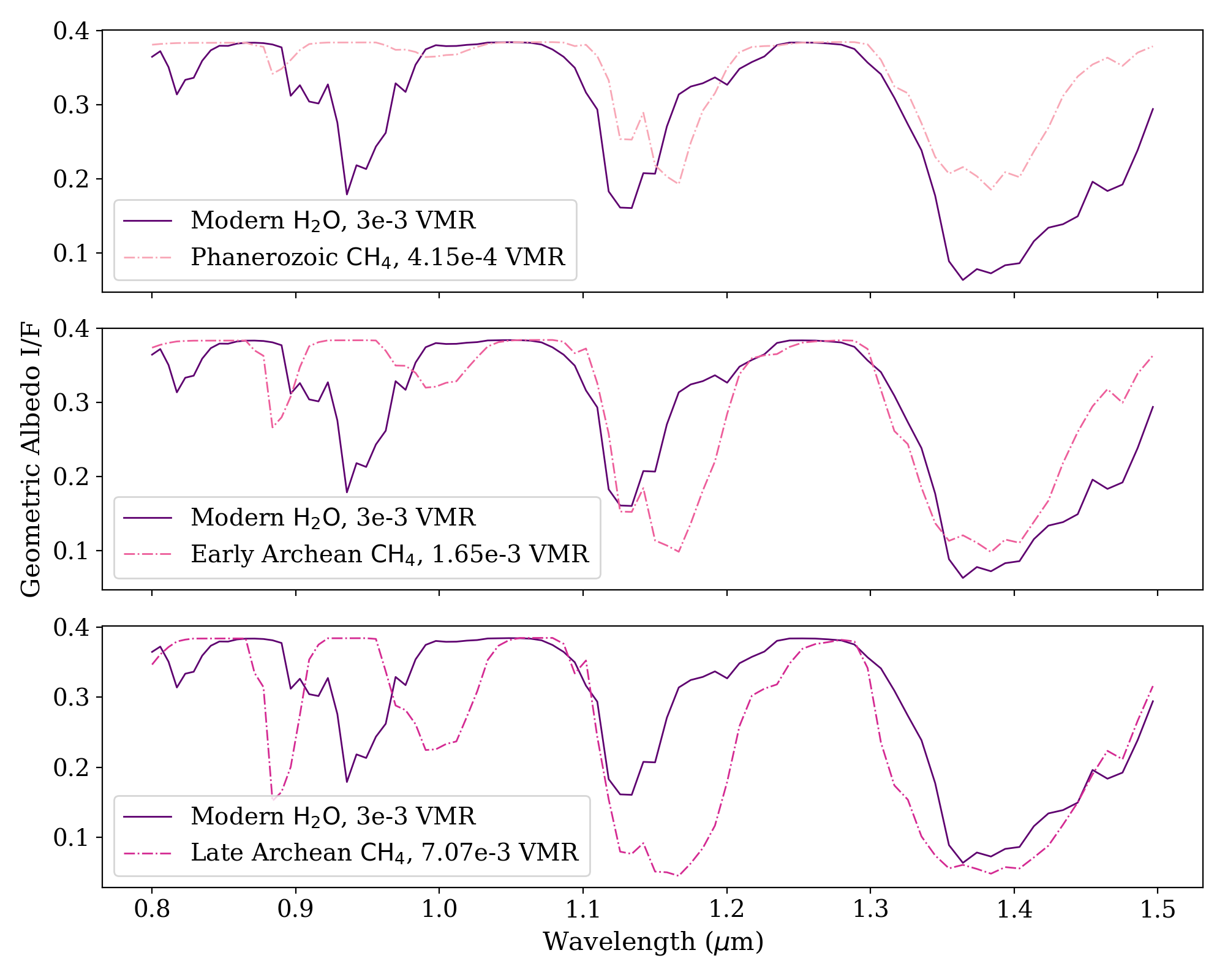}
\caption{Multi-panel plot illustrating overlap of \ce{H2O} and \ce{CH4} features in the NIR wavelength regime. Wavelength is on the x-axis, Geometric Albedo (I/F) is on the y-axis. Each panel shows a modern level of \ce{H2O} in dark purple, and a Phanerozoic, Early Archean, and Late Archean abundances of \ce{CH4} in the top, middle, and bottom panels with a dot-dashed shade of pink, respectively.}
\label{fig:varych4modernh2o}
\end{figure*}
\subsection{\ce{H2O} and \ce{CH4} Detection Degeneracy}

We ran a suite of retrievals with all combined \ce{H2O} and \ce{CH4} values shown in Table~\ref{tab:h2och4combo}, at 20\%, 30\%, and 40\% bandpass widths - note that these values were selected purely to understand the inflection point of molecular abundance and detectability, with the highest \ce{H2O} abundance at modern levels (3$\times10^{-3}$ VMR) and the highest \ce{CH4} abundance at Archean levels (7.07$\times10^{-3}$ VMR). Our lowest tested value of \ce{CH4} is (1$\times10^{-4}$ VMR), the first abundance at which any detection is possible. 

\begin{table}[h!]
\begin{centering}
\begin{tabular}{cc}
            \hline
            \hline
            \textbf{\ce{CH4} (VMR)} & \textbf{\ce{H2O} (VMR)} \\
            \hline
            3$\times10^{-5}$ & 3$\times10^{-5}$ \\
            1$\times10^{-4}$ & 1$\times10^{-4}$ \\
            3$\times10^{-4}$ & 3$\times10^{-3}$ \\
            1$\times10^{-3}$ & 1$\times10^{-4}$ \\
            3$\times10^{-3}$ & 3$\times10^{-3}$ \\
            7$\times10^{-3}$ & 1$\times10^{-2}$\\
            \hline
\end{tabular}
\caption{All values for \ce{CH4} and \ce{H2O} run for our analysis of detectability dependence on molecular abundance. We tested all combinations of abundances, leading to 25 abundance combinations.}
\label{tab:h2och4combo}
\end{centering}
\end{table}

In Figure~\ref{fig:heatmapsabundancech4}, we present the results of our degeneracy study as six heatmaps at 1.1 {\microns} and the longest possible bandpass (top and bottom rows, respectively), wherein the color corresponds to the SNR, with the required SNR for strong detection labeled on the plot. The x-axis is the \ce{H2O} abundance, the y-axis is \ce{CH4} abundance, with a 20\%, 30\%, and 40\% bandpass width, respectively, in columns. All results are with R=70, the current fiducial resolving power in the NIR. We can see as a trend in Figure~\ref{fig:heatmapsabundancech4} that the bandpass width has a very large impact on detectability for \ce{CH4} as a function of abundances: at high \ce{CH4} abundances, especially at 1.1 {\microns} where there is a large \ce{CH4} absorption feature, Archean levels of \ce{CH4} remain essentially unchanged through bandpass width. This reaffirms our earlier results that high \ce{CH4} did not vary through bandpass width and that \ce{H2O} abundance is a large factor in detectability. Looking first to Figures~\ref{fig:heatmapsabundancech4}a - \ref{fig:heatmapsabundancech4}c, at high \ce{CH4} values, 3$\times10^{-3}$ VMR and above, the \ce{H2O} abundance minimally influences the \ce{CH4} detectability. We begin to see an influence in detectability at 1$\times10^{-3}$ VMR, where at the lowest \ce{H2O} abundance (3$\times10^{-5}$ VMR) the required SNR for \ce{CH4} detection is $sim$4, through the central \ce{H2O} abundances, the required SNR is $sim$6 for \ce{CH4} detection, and at the highest \ce{H2O} abundance (1$\times10^{-2}$ VMR), the required SNR for \ce{CH4} detection is up to 11 at a 20\% bandpass, and an 8 for a 40\% bandpass. The \ce{CH4} abundance is unchanging across this range of \ce{H2O} abundances; thus it is clear that the influence must be \ce{H2O}. This influence continues to grow stronger as the \ce{CH4} abundance decreases, with 3$\times10^{-4}$ VMR \ce{CH4} abundance requiring an SNR of 8 at the lowest \ce{H2O} abundances for a 20\% bandpass and a 6 for 30\% and 40\%, and an SNR higher than 20 (i.e., undetectable in our study) at the highest \ce{H2O} abundances. The degeneracy is most obvious at 1$\times10^{-4}$ VMR \ce{CH4} abundance, with each increasing abundance of \ce{H2O} increasing the required SNR for strong \ce{CH4} detection - and resulting in no detections. 

In Figures~\ref{fig:heatmapsabundancech4}d - \ref{fig:heatmapsabundancech4}f, we adjust the wavelength that is being investigated. Since the bandpass centers change as the bandpass width increases, the final bandpass is investigated in \ref{fig:heatmapsabundancech4}d - \ref{fig:heatmapsabundancech4}f with different bandpass centers. The emphasis in this last bandpass is in the 1.3 {\microns} feature that appears for both \ce{CH4} and \ce{H2O}. We find a  similar trend to the above, with higher SNR requirements at each bandpass. With the 20\% bandpass width (Figure~\ref{fig:heatmapsabundancech4}d) specifically, the required SNR is higher than in Figures~\ref{fig:heatmapsabundancech4}a - \ref{fig:heatmapsabundancech4}c, especially at high \ce{H2O}. In fact, at the highest value of \ce{H2O} (1$\times10^{-2}$ VMR), \ce{CH4} is not detectable at 3$\times10^{-3}$ VMR and 1$\times10^{-3}$ VMR although both of those abundances were still detectable at shorter wavelengths. Thus longer wavelengths (e.g. 1.35 {\microns}) are not well suited for \ce{CH4} detection, especially with narrower bandpasses.

Next, we investigate the \ce{H2O} detectability. We present in Figure~\ref{fig:heatmapsabundanceh2o} another set of heatmaps to investigate how the detection of \ce{H2O} changes as a function of \ce{CH4} abundance. All plot aspects remain the same as Figure~\ref{fig:heatmapsabundancech4}, with the exception that we are investigating \ce{H2O} detectability. We again see that there is a strong correlation with the bandpass width, with detectability increasing as the bandpass width increases, especially at shorter wavelengths. At shorter wavelengths, such as 1.1 {\microns}, \ce{H2O} is detectable at all but the lowest abundances, with changing SNR. At a modern-Earth level of \ce{H2O}, detectability at a 20\% bandpass width requires at most an SNR of 12 and more than halves to an SNR$\le$5 for a bandpass width of 40\%. However, this does not hold for long wavelengths. We find that at longer wavelengths, the high \ce{CH4} abundances completely mask the \ce{H2O} features leading to an inability to detect \ce{H2O} at high \ce{CH4} or low \ce{H2O}. However, we did find that at low \ce{H2O} and sufficiently low \ce{CH4}, the presence of both molecules together increases the detectability of \ce{H2O}, likely due to the degeneracy between the two molecules being broken in this wavelength regime.

\begin{figure*}[ht!]
\centering
\gridline{\fig{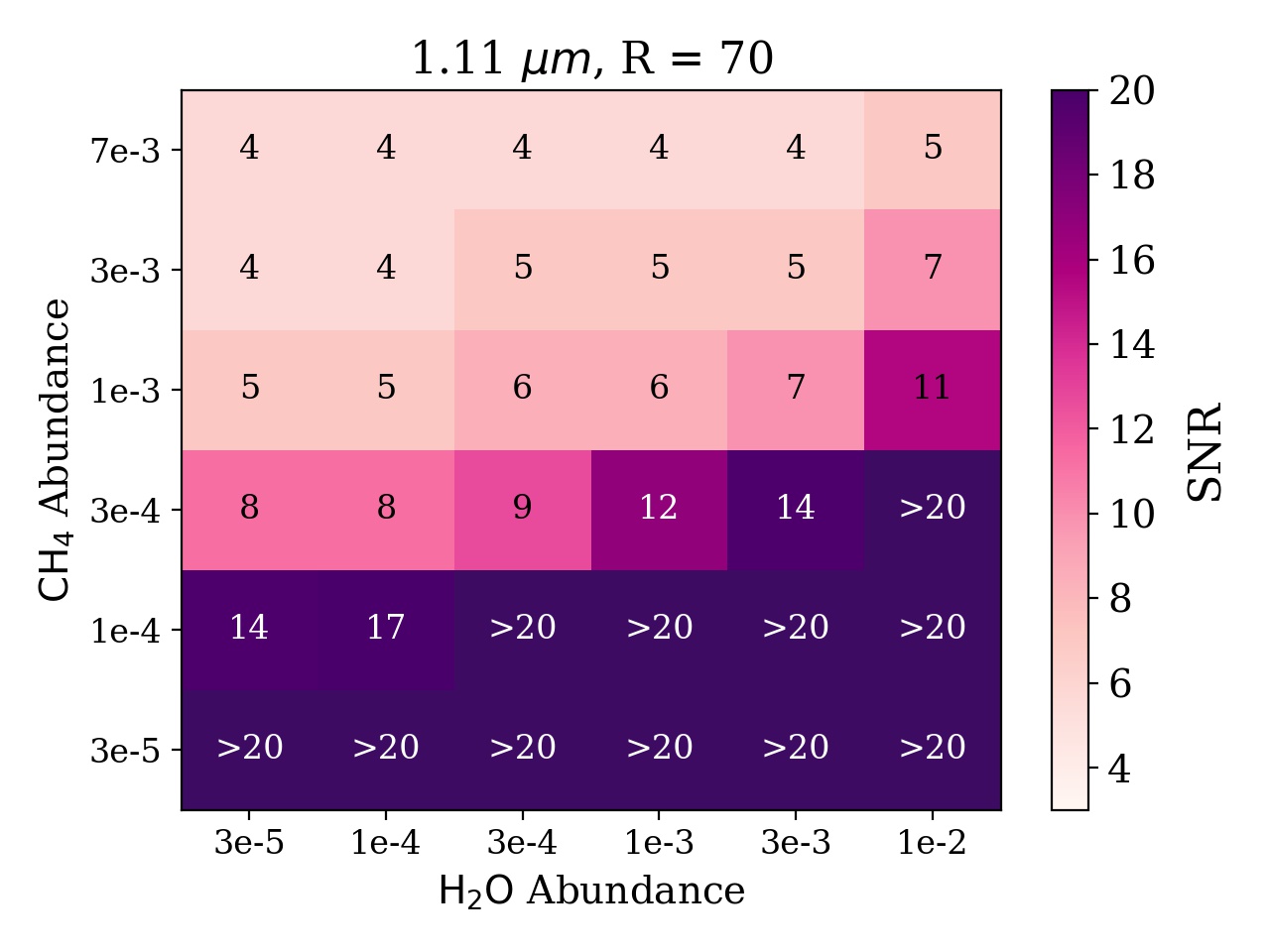}{0.334\textwidth}{\vspace{-0.35cm}(a) 20\% bandpass, 1.008 -- 1.229 {\microns}}
          \fig{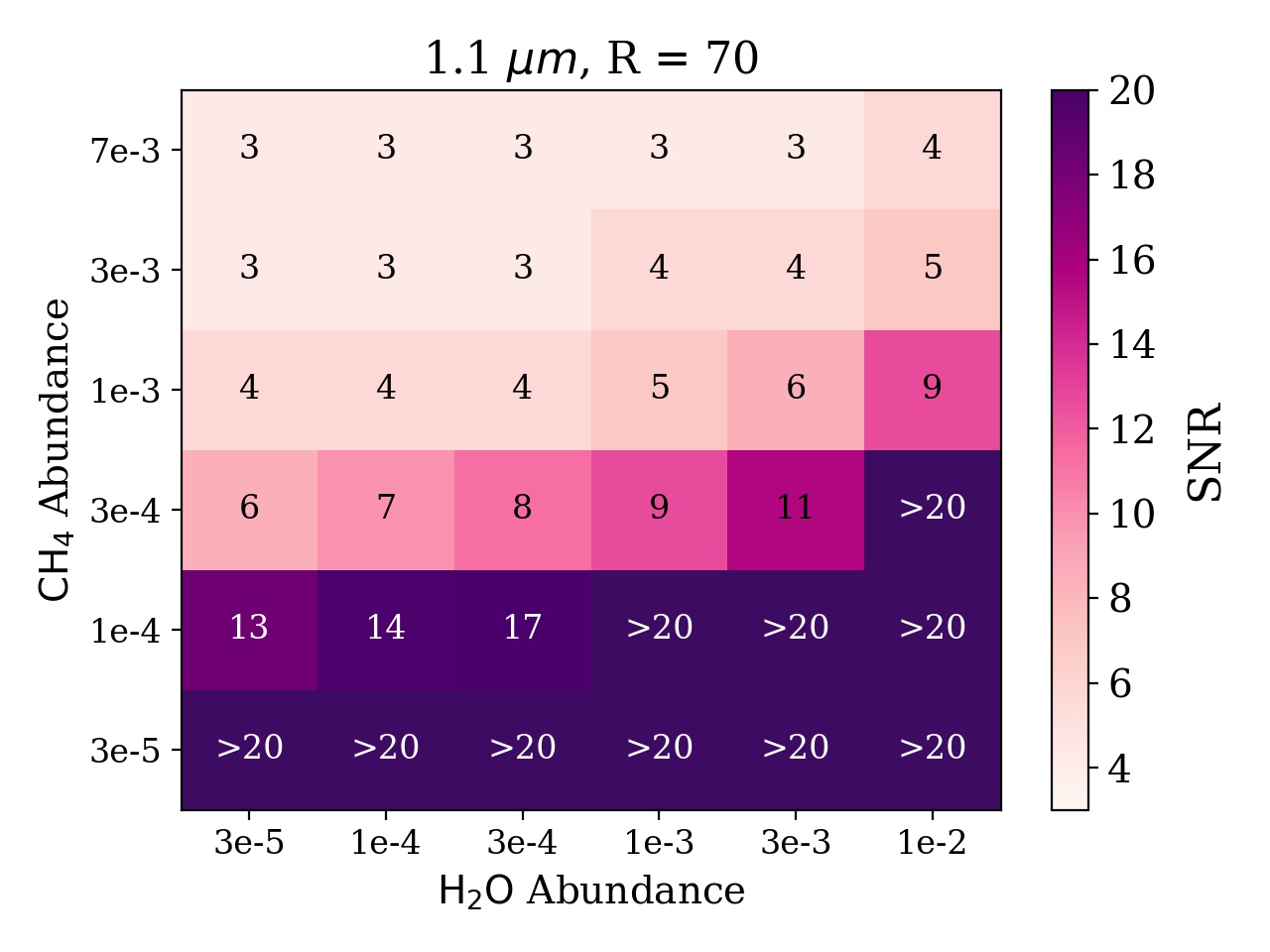}{0.334\textwidth}{\vspace{-0.35cm}(b) 30\% bandpass, 0.952 -- 1.282 {\microns}}
          \fig{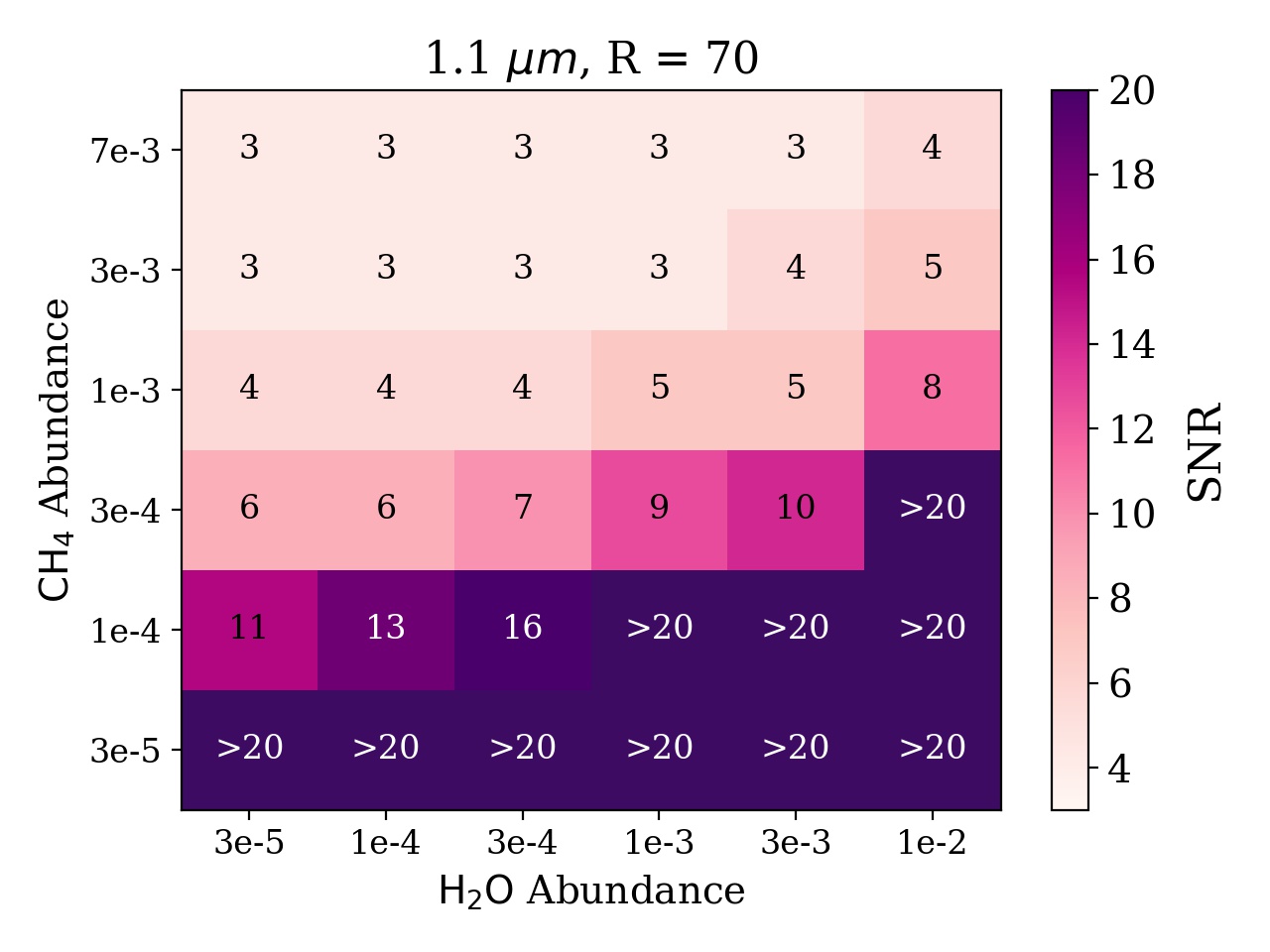}{0.334\textwidth}{\vspace{-0.35cm}(c) 40\% bandpass, 0.912 -- 1.338 {\microns}}}
\vspace{-0.4cm}
\gridline{\fig{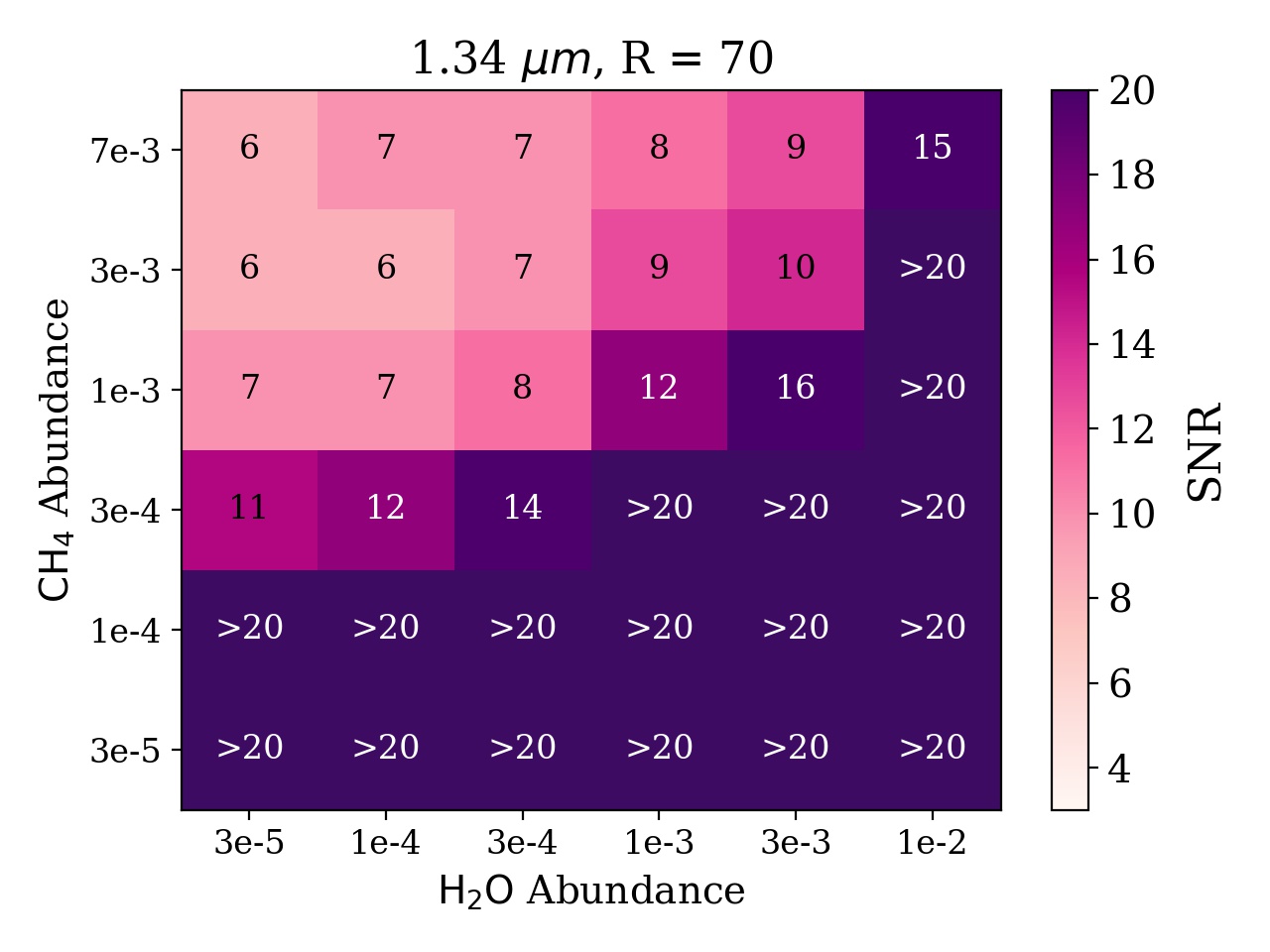}{0.334\textwidth}{\vspace{-0.35cm}(d) 20\% bandpass, 1.229 -- 1.478 {\microns}}
          \fig{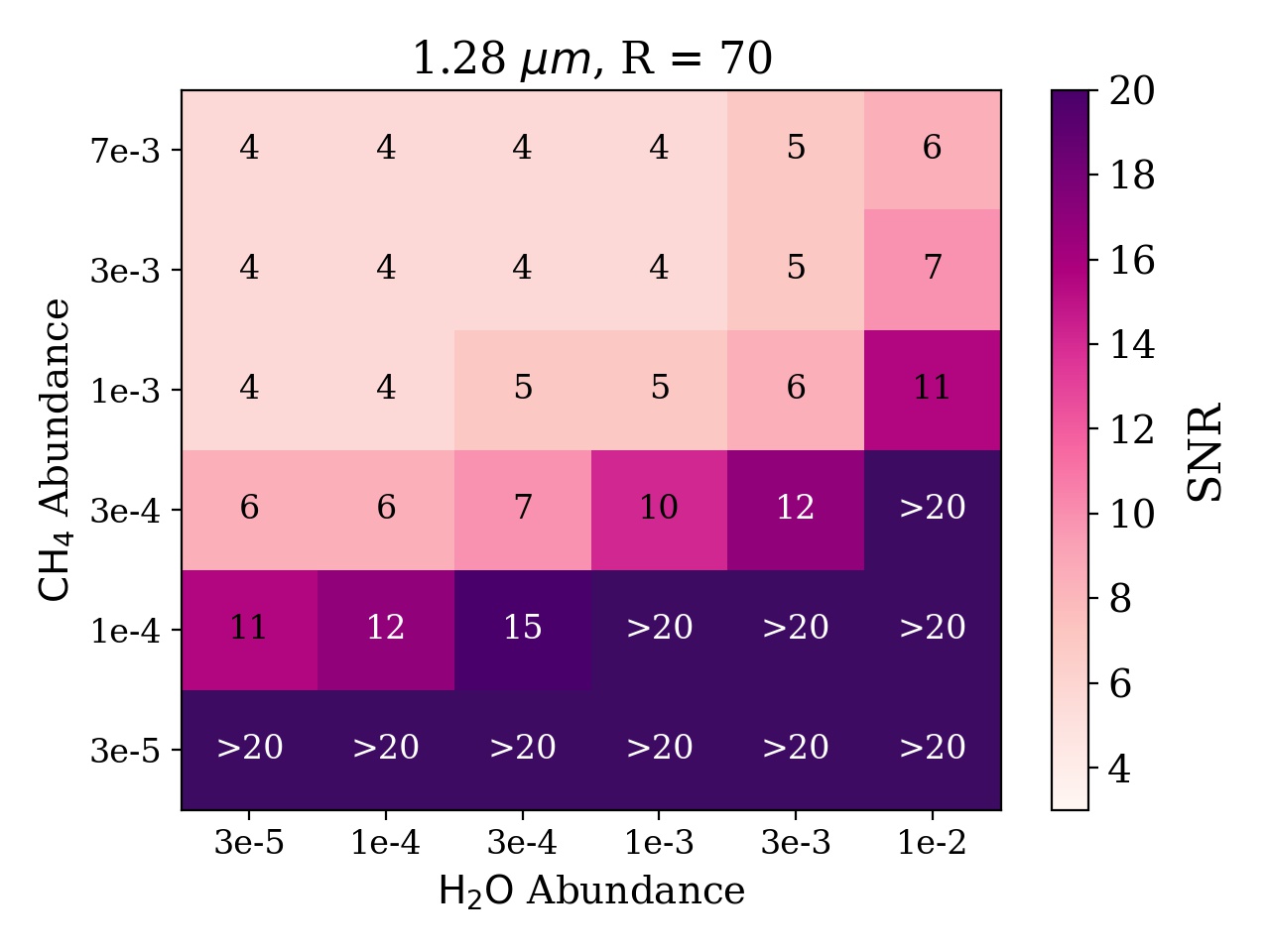}{0.334\textwidth}{\vspace{-0.35cm}(e) 30\% bandpass, 1.113 -- 1.478 {\microns}}
          \fig{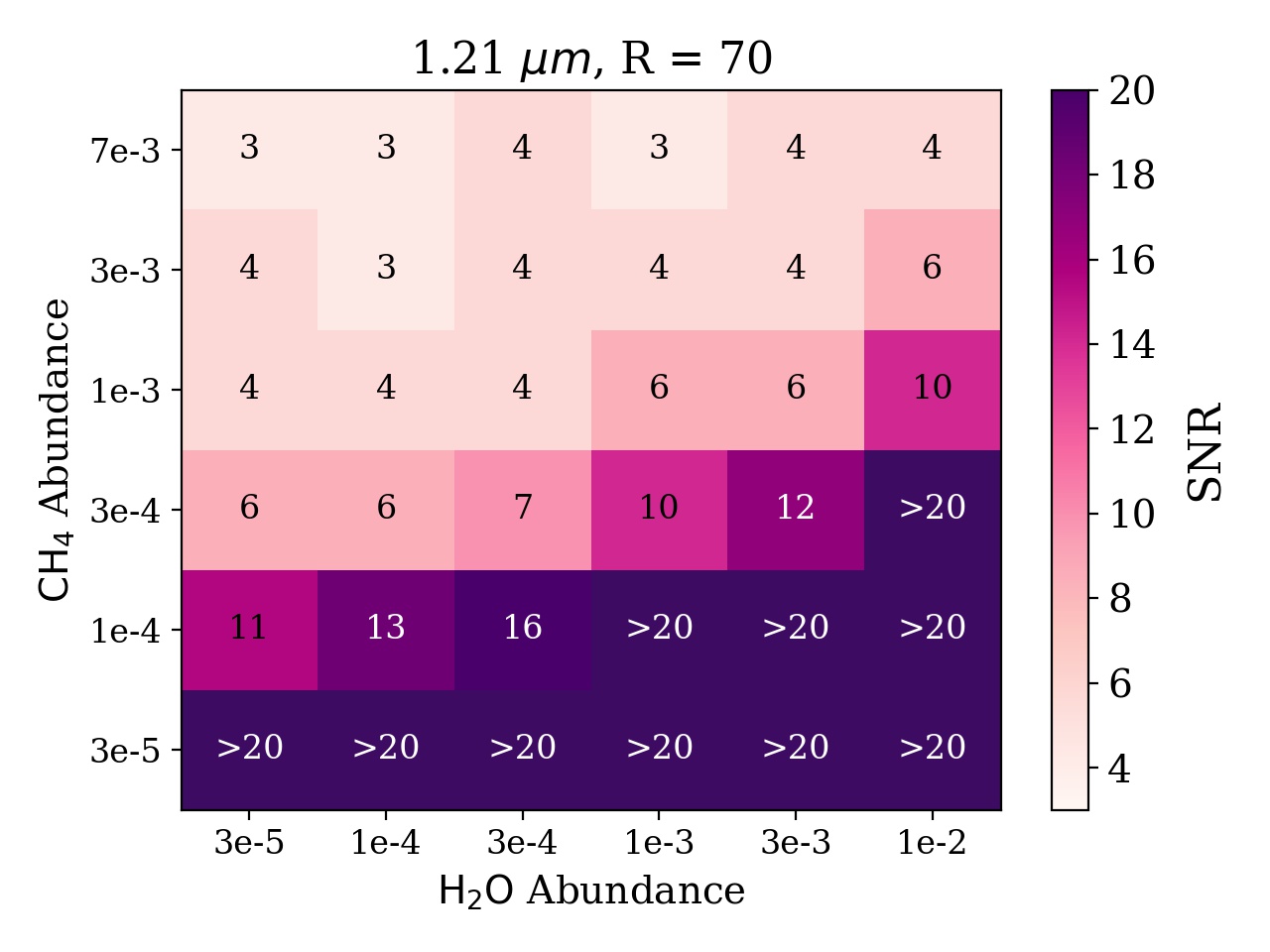}{0.334\textwidth}{\vspace{-0.35cm}(f) 40\% bandpass, 0.993 -- 1.478 {\microns}}}
\vspace{-0.2cm}
\caption{\textbf{SNRs for Strong \ce{CH4} Detection.} Heatmaps of \ce{CH4} detectability as a function of molecular abundance for R = 70, at 1.11 {\microns} (top row) and centered on the 1.3 {\microns} feature (bottom row). We present 20\%, 30\%, and 40\% bandpass widths. y-axis: \ce{CH4} abundance, x-axis:\ce{H2O} abundance, heat: SNR. Required SNRs for strong detection are written on the associated block, unless there is no strong detection, thus labeled `$>$20'.}
\label{fig:heatmapsabundancech4}
\end{figure*}

\begin{figure*}
\centering
\gridline{\fig{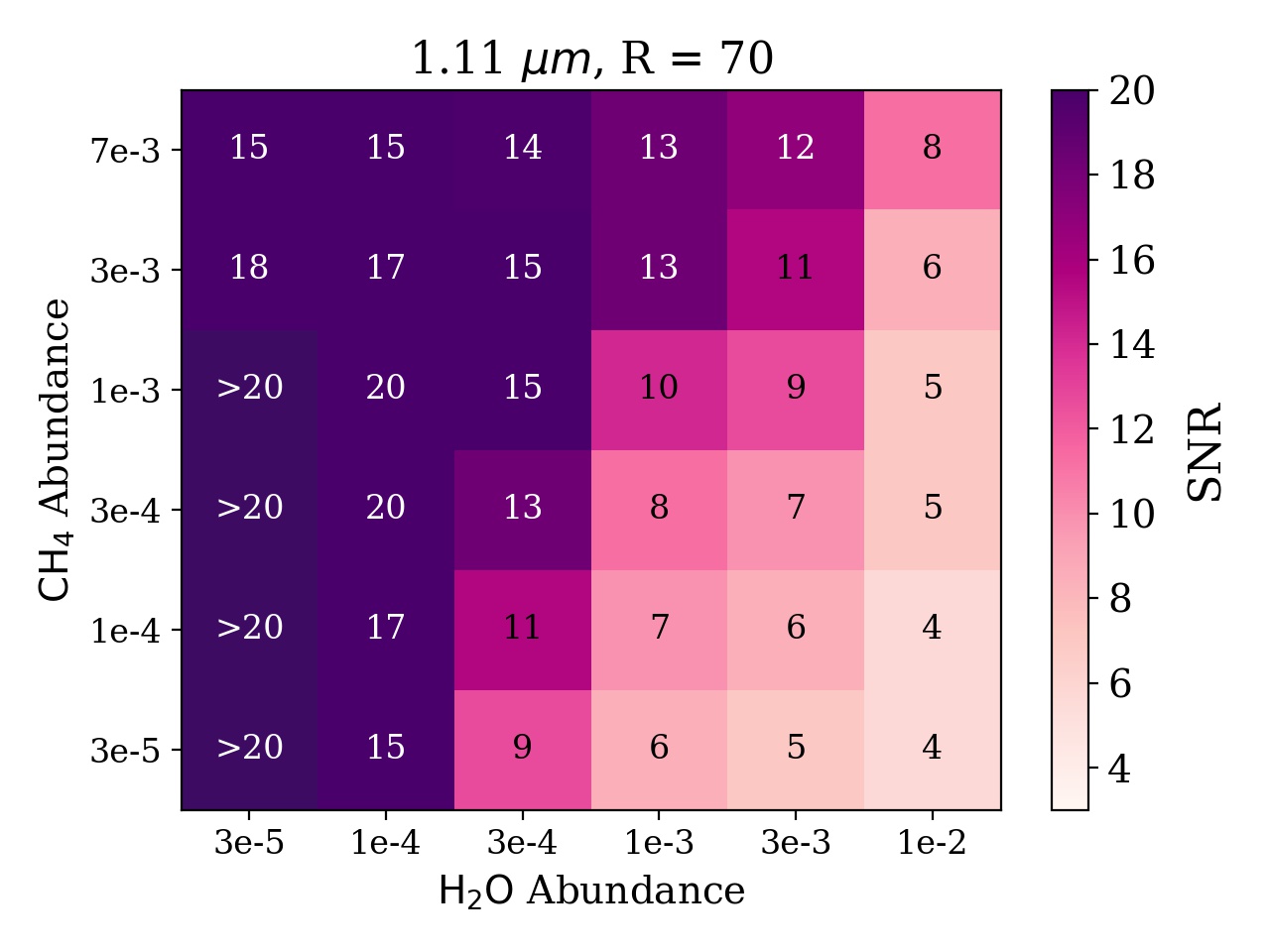}{0.334\textwidth}{\vspace{-0.35cm}(a) 20\% bandpass, 1.008 -- 1.229 {\microns}}
          \fig{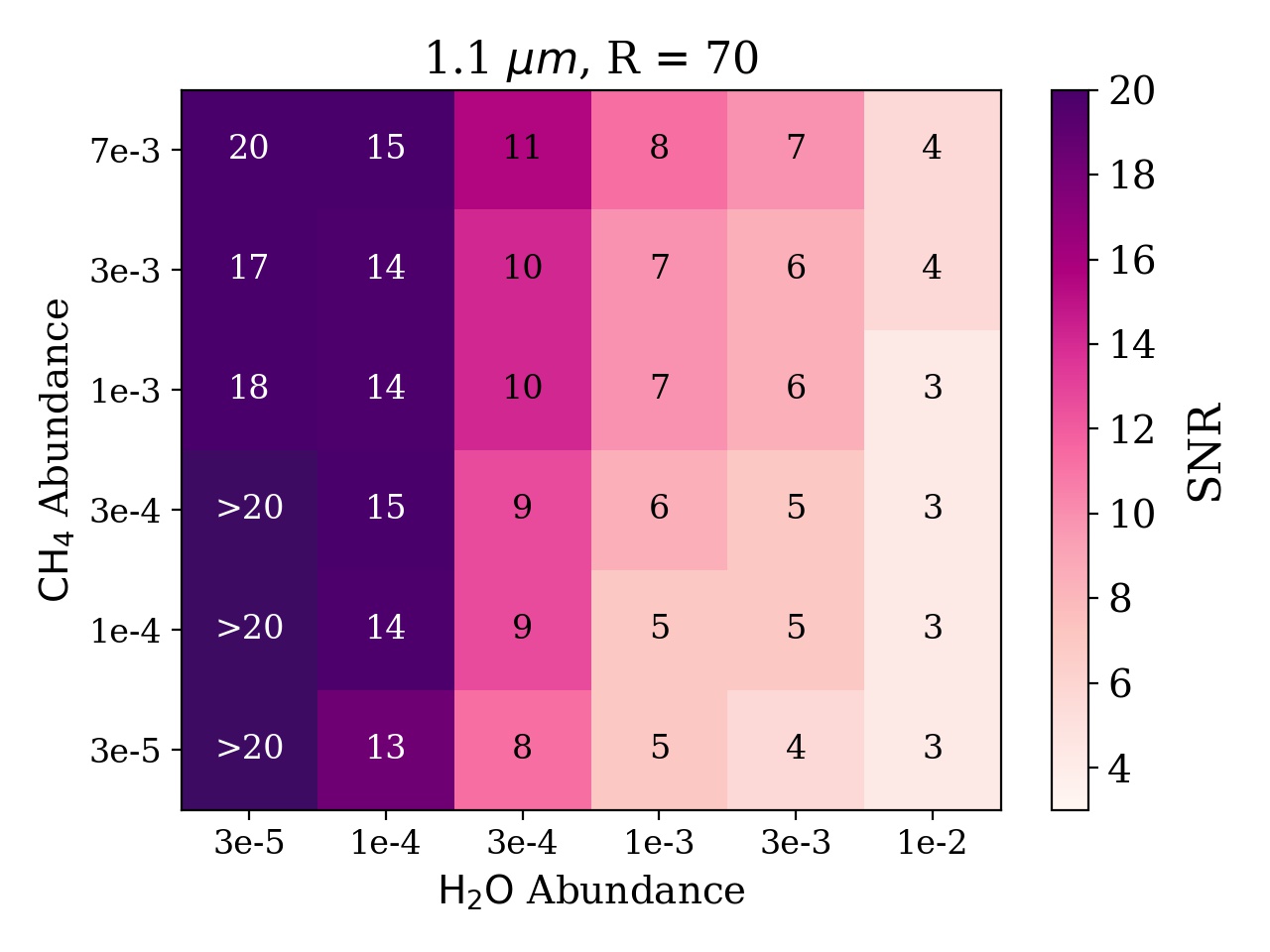}{0.334\textwidth}{\vspace{-0.35cm}(b) 30\% bandpass, 0.952 -- 1.282 {\microns}}
          \fig{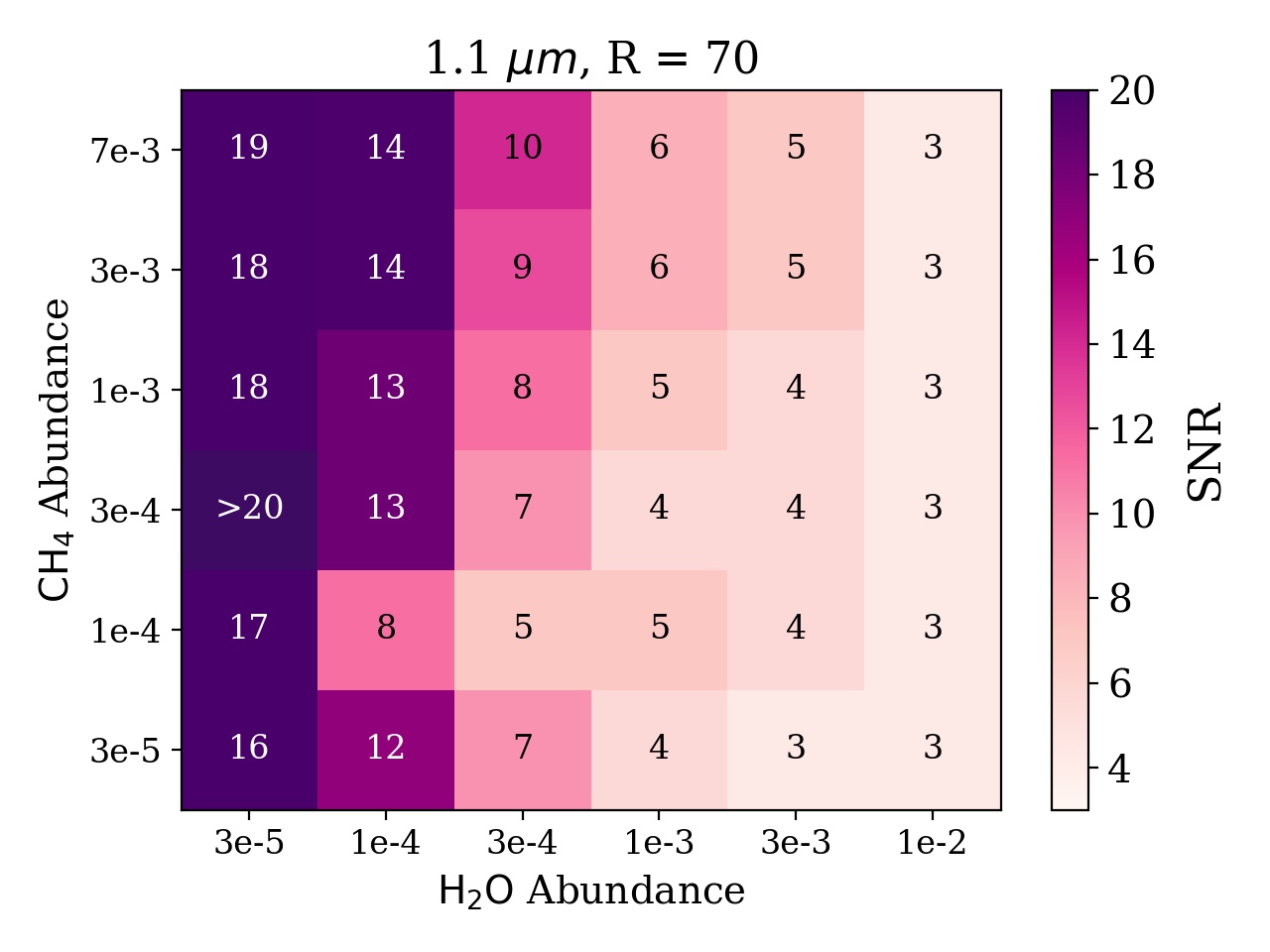}{0.334\textwidth}{\vspace{-0.35cm}(c) 40\% bandpass, 0.912 -- 1.338 {\microns}}}
\vspace{-0.4cm}
\gridline{\fig{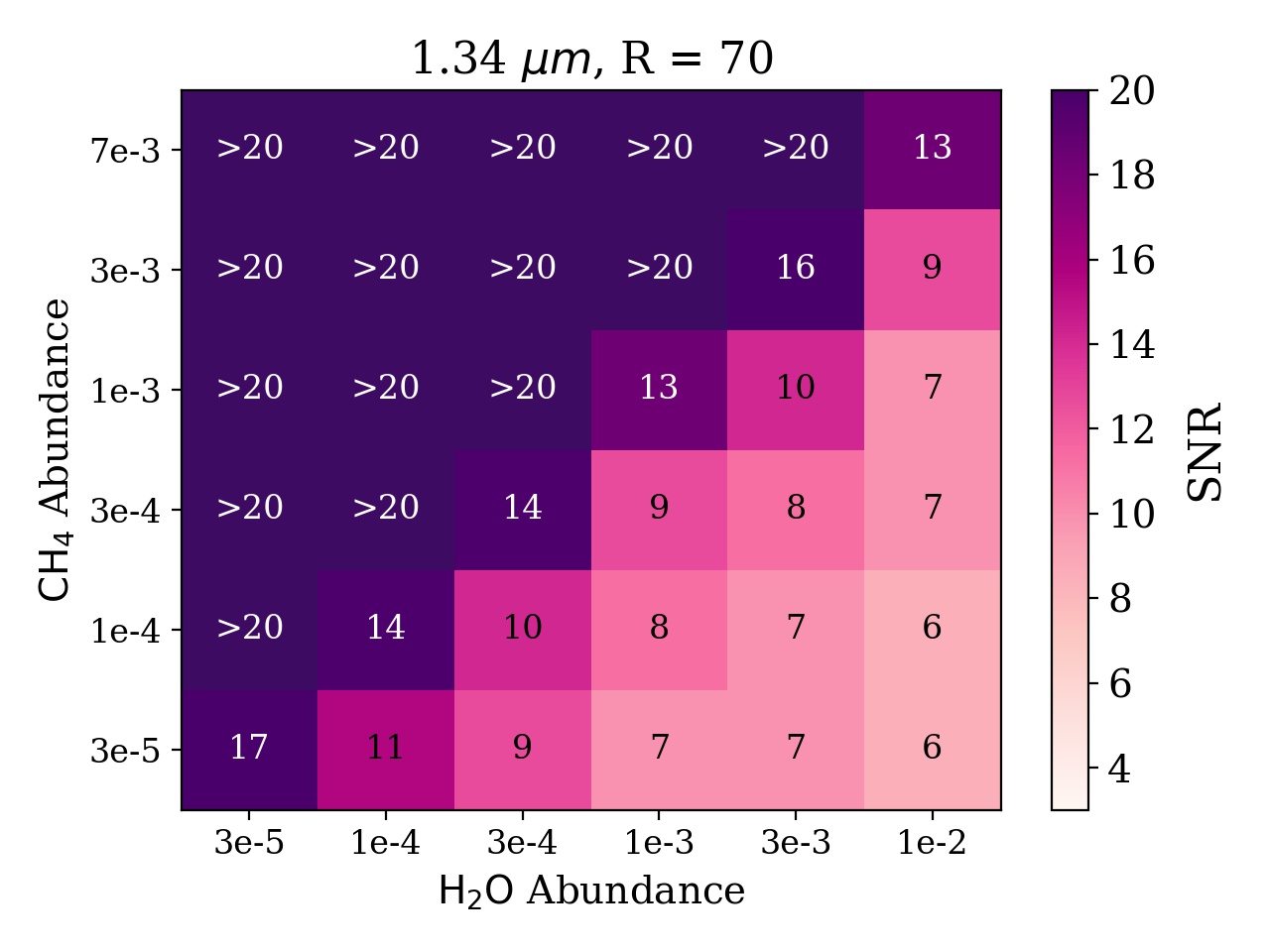}{0.334\textwidth}{\vspace{-0.35cm}(d) 20\% bandpass, 1.229 -- 1.47 8{\microns}}
          \fig{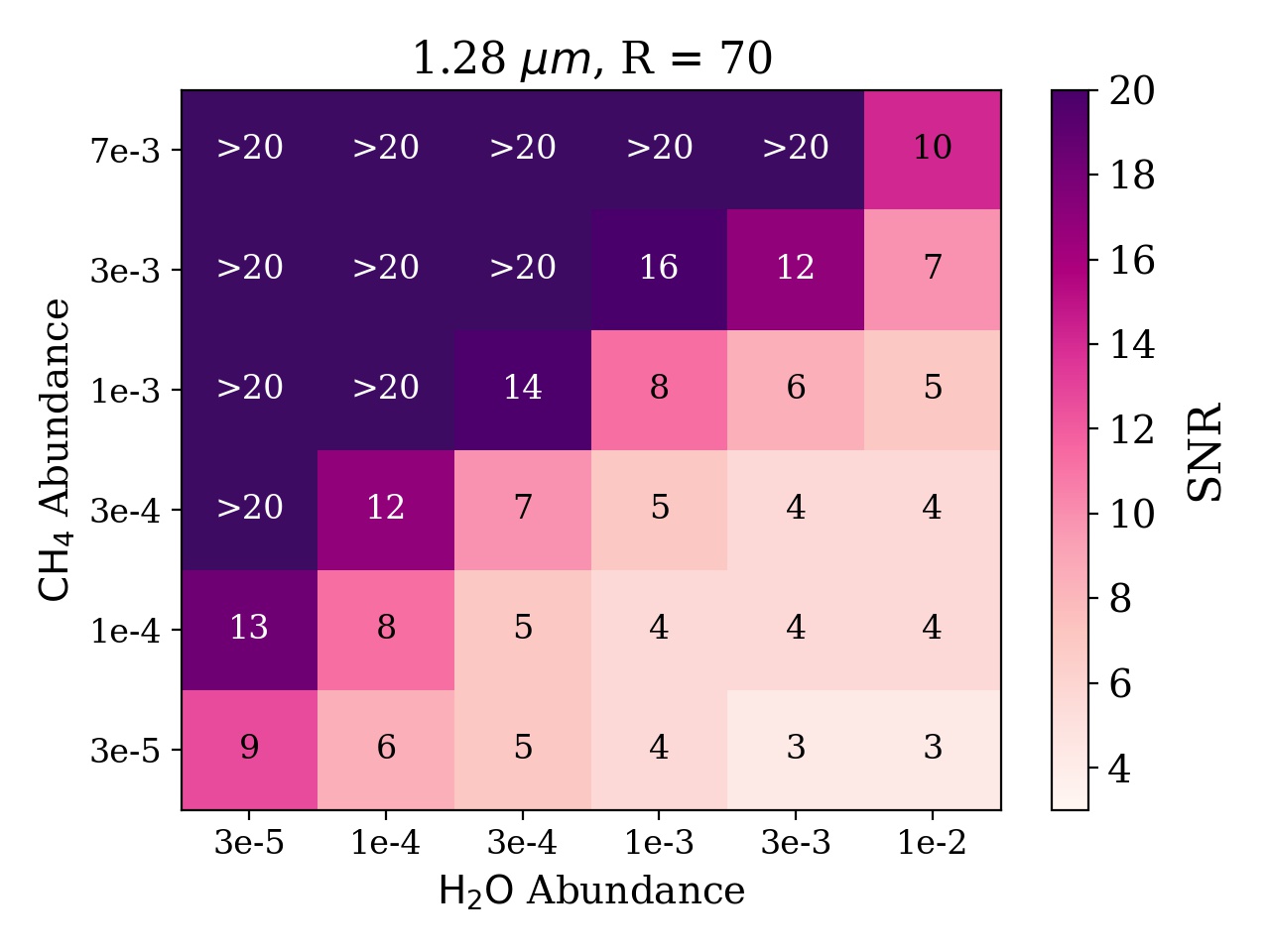}{0.334\textwidth}{\vspace{-0.35cm}(e) 30\% bandpass, 1.113 -- 1.478 {\microns}}
          \fig{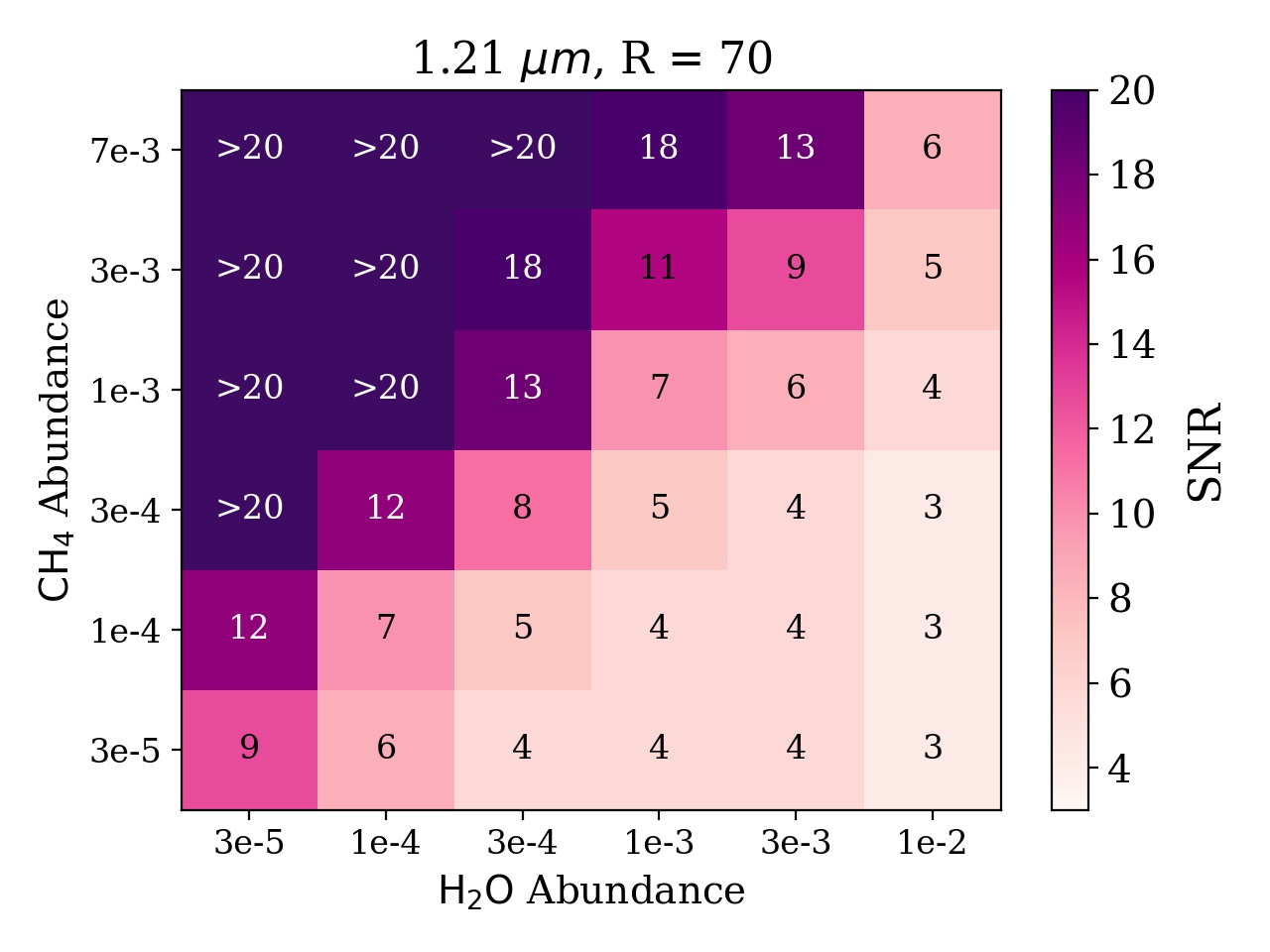}{0.334\textwidth}{\vspace{-0.35cm}(f) 40\% bandpass, 0.993 -- 1.478 {\microns}}}
\vspace{-0.2cm}
\caption{\textbf{SNRs for Strong \ce{H2O} Detection.} All plot facets remain the same as Figure~\ref{fig:heatmapsabundancech4}, but illustrating \ce{H2O} detectability.}
\label{fig:heatmapsabundanceh2o}
\end{figure*}

\section{Discussion}
\label{sec:disc}

Understanding the potential for dual detection of atmospheric constituents is an important factor in optimizing the required telescope observing time. Figure~\ref{fig:h2och4}a shows how crucial this investigation is in the NIR. At the lowest detectable values of \ce{CH4} in our initial investigation, in order to detect both \ce{H2O} and \ce{CH4} an SNR of 9 is required at 1.1 {\microns}, and this remains possible out to 1.2 {\microns}. In order to detect both \ce{H2O} and low \ce{CH4} at shorter wavelengths, such as at 0.9 {\microns}, an SNR of $\geq$13 is required. This is a significant difference in SNR and would result in a drastic increase in required exposure time. However, the optimal wavelength for a detection of both molecules reverses at higher \ce{CH4} abundances - the detectability of \ce{H2O} is impacted at longer wavelengths as shown in Figure~\ref{fig:h2och4}c. In order to detect both \ce{H2O} and \ce{CH4} at 1.1 {\microns}, an SNR of $\geq$15 is required; however, at 0.9 {\microns} the combined detection is accessible at an SNR of 5. We begin to see that there is a relationship between \ce{H2O} and \ce{CH4}, which is further confirmed by the degeneracy between \ce{H2O} and \ce{CH4} shown in Figure~\ref{fig:archeancorner}.

The overlap of molecular absorption features as shown in Figure~\ref{fig:varych4modernh2o} drives this degeneracy. The bottom most panel shows both a modern Earth \ce{H2O} abundance (3$\times10^{-3}$ VMR) and a late Archean Earth \ce{CH4} abundance (7.07$\times10^{-3}$ VMR). At 0.9 {\microns}, we can see the absorption features are orthogonal to each other and thus more easily disentangled. At 1.1 {\microns} and longer, the \ce{CH4} feature absorbs much deeper and wider than the \ce{H2O} feature, thus completely preventing the possibility of detection without a high SNR. The opposite effect is seen at lower \ce{CH4} abundance, such as the Phanerozoic abundance (4.15$\times10^{-4}$ VMR) shown in the top panel of Figure~\ref{fig:varych4modernh2o}. Here, the \ce{CH4} features at 0.9 {\microns} are too shallow to be detected without high SNR, and at longer wavelengths (such as 1.4 {\microns}), the \ce{CH4} feature is completely contained within a much larger \ce{H2O} feature and is thus not detectable. At 1.1 {\microns}, both \ce{CH4} and \ce{H2O} are detectable due to the slight offset between the absorption features. We can clearly see that the detectability is a function of the abundance of \ce{CH4} and \ce{H2O}, and the abundances of \ce{CH4} and \ce{H2O} are degenerate in certain regimes.

To further investigate this degeneracy, we first looked at comparing a null versus modern \ce{H2O} value in Figure~\ref{fig:heatmapsnowater}, specifically at 1.1 {\microns} where the overlapping features compete. In Figure~\ref{fig:heatmapsnowater}a with modern Earth \ce{H2O} abundance, we achieve a strong \ce{CH4} detection at 4.15$\times10^{-4}$ VMR at SNR = 9, while still achieving at most a weak detection at 1$\times10^{-4}$. Comparatively, in Figure~\ref{fig:heatmapsnowater}b 4.15$\times10^{-4}$ VMR requires an SNR of 6 for strong detection with no \ce{H2O} present, compared to the previous SNR of 9. This confirms thatd \ce{H2O} are degenerate, with detectability reliant on the abundances, and thus we explored a range of abundance values of \ce{H2O} and \ce{CH4} to finely understand the relationship we found.

In Figures~\ref{fig:heatmapsabundancech4} and \ref{fig:heatmapsabundanceh2o}, we see that there are clear degeneracies between \ce{H2O} and \ce{CH4}, but with further varying relationships through wavelength. We find that the detectabilities of \ce{H2O} and \ce{CH4} are both a function of \ce{CH4} and \ce{H2O} abundances, wavelength, and bandpass width. In Figures~\ref{fig:heatmapsabundancech4}a - \ref{fig:heatmapsabundancech4}c and \ref{fig:heatmapsabundanceh2o}a - \ref{fig:heatmapsabundanceh2o}c, we can clearly see that both molecules are far easier to detect, at all bandpass widths, in this wavelength regime when comparing to Figures~\ref{fig:heatmapsabundancech4}d - \ref{fig:heatmapsabundancech4}f and \ref{fig:heatmapsabundanceh2o}d - \ref{fig:heatmapsabundanceh2o}f presenting \ce{CH4} and \ce{H2O} at longer wavelengths. At longer wavelengths, both molecules are more difficult to detect, with more abundance values undetectable. Critically, a modern Earth level of \ce{H2O} is detectable at all \ce{CH4} abundances and bandpass widths at 1.1 {\microns} while it is undetectable at 20\% and 30\% bandpasses at Archean levels of \ce{CH4} at 1.3 {\microns} . Similarly, \ce{CH4} is more accessibly detectable across bandwidth at shorter wavelengths, and detectablity worsens across the board at longer wavelengths. With this, we can dismiss the possibility of looking for \ce{CH4} and \ce{H2O} at the longer wavelengths considered in this study (e.g. 1.35 {\microns} -- 1.5  {\microns}. 

\begin{figure*}
\centering
\gridline{
          \fig{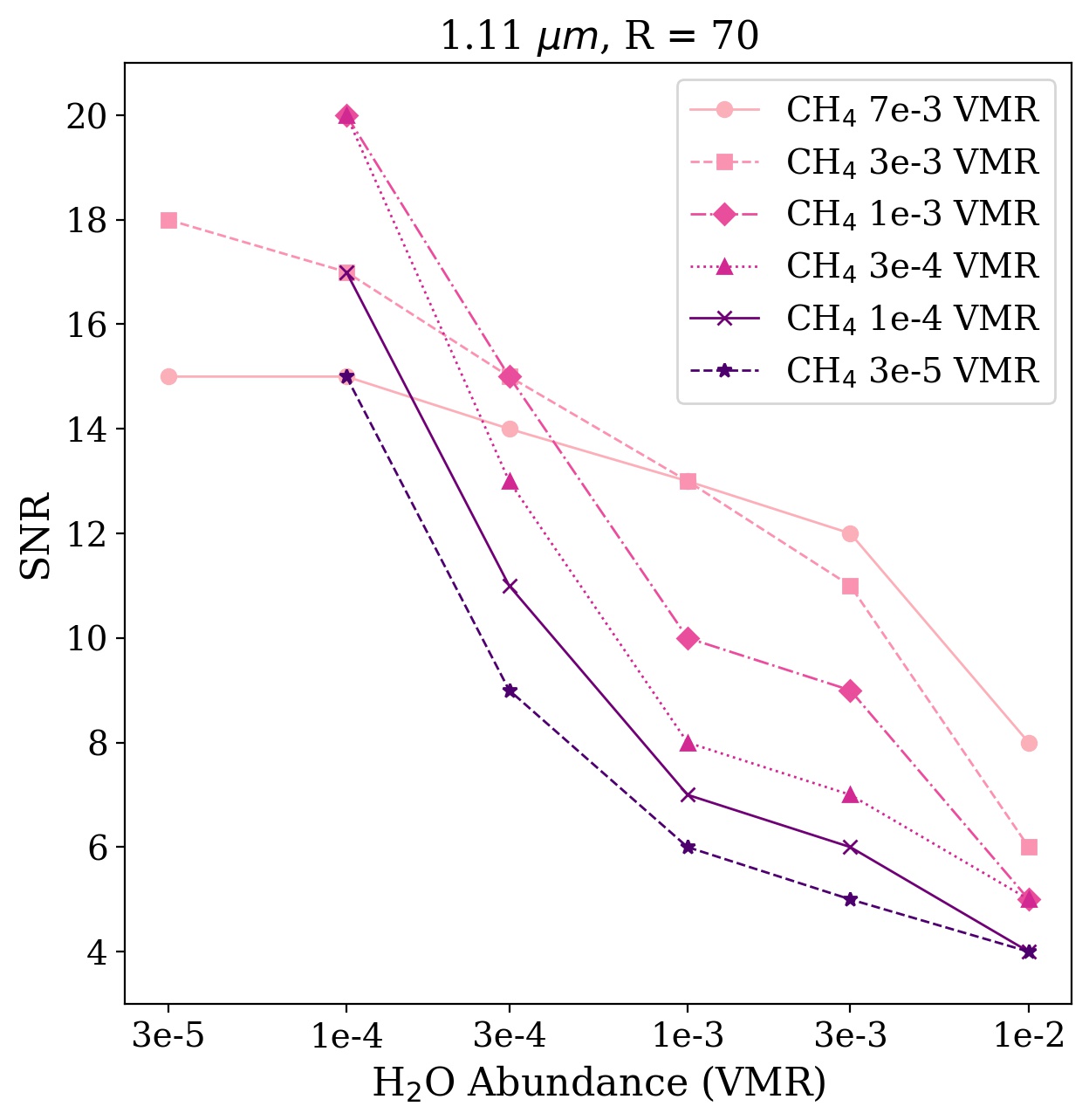}{0.45\textwidth}{(a) 20\% bandpass, 1.008 -- 1.229 {\microns}}
          \fig{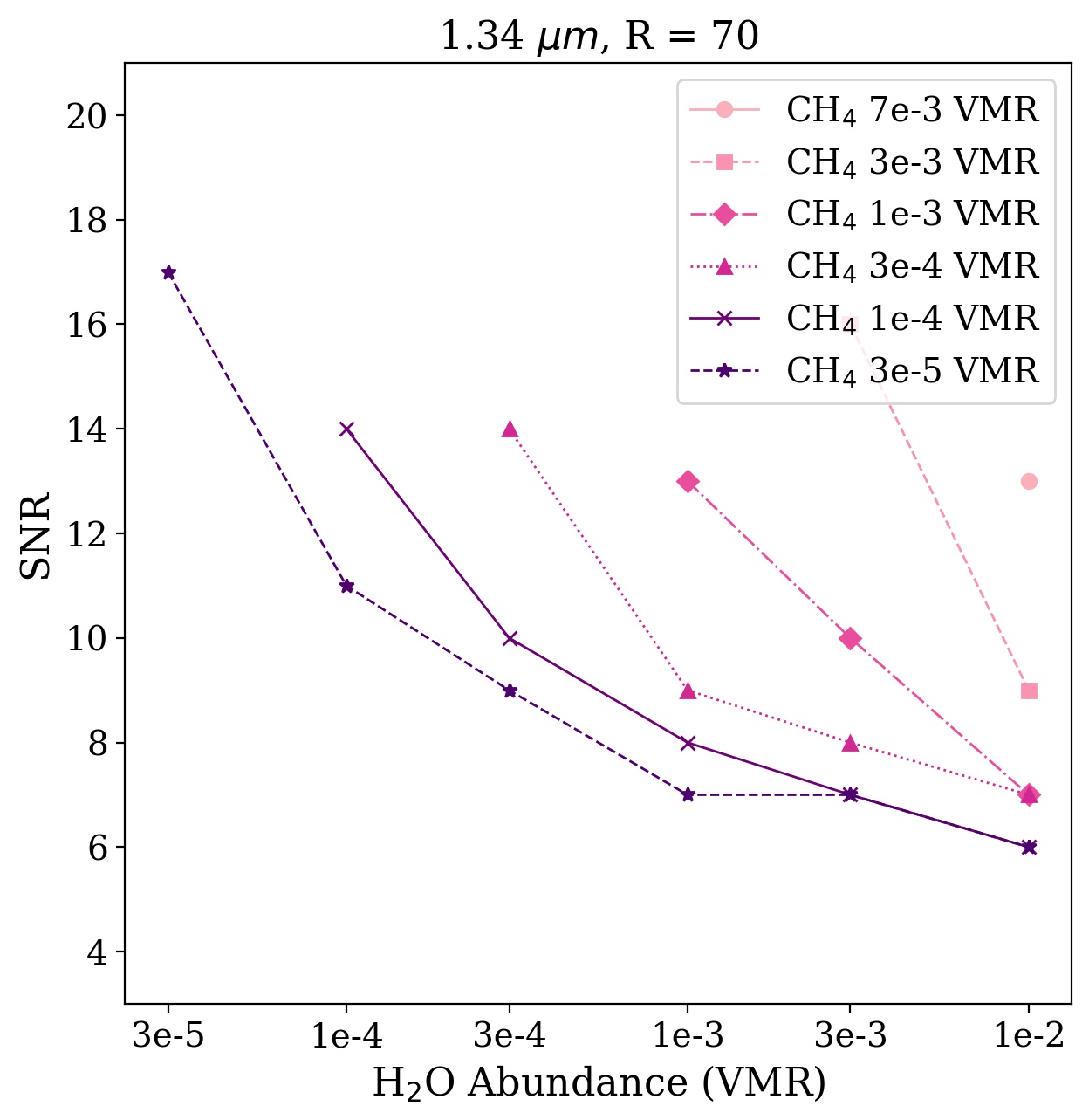}{0.45\textwidth}{(b) 20\% bandpass, 1.229 -- 1.478 {\microns}}}
\caption{\textbf{SNRs for Strong \ce{H2O} Detection.} Plot of the lowest SNR values at which a strong detection is achieved for all \ce{H2O} values in our degeneracy investigation at (a) 1.1 {\microns} and (b) 1.34 {\microns} with a 20\% bandpass at R=70. The \ce{H2O} VMR values are on the x-axis, with SNR on the y-axis. Each line represents a different tested \ce{CH4} value in VMR, labeled in the figure with different line styles and markers.}
\label{fig:lineh2odetec}
\end{figure*}

Focusing on the shorter, 1.1 {\microns} wavelength (i.e., Figures~\ref{fig:heatmapsabundancech4}a - \ref{fig:heatmapsabundancech4}c), we see a steady trend of increased SNR requirements for \ce{CH4} abundances as a function of \ce{H2O} abundance. At the Archean abundance, which is of high interest, we see that there is essentially no impact to the \ce{CH4} SNR requirements. However at lower values that can be used as guides for other epochs, such as 3$\times10^{-4}$ VMR for the Phanerozoic epoch, we see changes in the required SNR for \ce{CH4} detection. At the lowest \ce{H2O} abundance an SNR of 8 is required at a 20\% bandpass width, while at the modern \ce{H2O} abundance (3$\times10^{-3}$ VMR) an SNR of 10 is required - thus the required SNR for detection almost doubles across \ce{H2O} abundance ranges, becoming undetectable at higher \ce{H2O} abundances. When moving to lower abundances of \ce{CH4} this effect is more pronounced, requiring higher base SNRs at the lowest \ce{H2O} abundance and becoming undetectable at mid to high \ce{H2O} abundances. This is especially crucial when we consider that, although a modern Earth abundance of \ce{CH4} (1.65$\times10^{-6}$ VMR) is not detectable, a high \ce{H2O} abundance can mask the detection of \ce{CH4} at low to mid-level values. 

Figure~\ref{fig:lineh2odetec} shows the \ce{CH4} and \ce{H2O} degeneracy in more detail. We present the minimum SNR required per \ce{H2O} abundance for \ce{H2O} detectability at each \ce{CH4} abundance at R=70 and two wavelengths: 1.1 {\microns} in Figure~\ref{fig:lineh2odetec}a and 1.34 in Figure~\ref{fig:lineh2odetec}b. In Figure~\ref{fig:lineh2odetec}a we can see that, as before, 3$\times10^{-5}$ VMR \ce{H2O} is only accessible at the two highest values of \ce{CH4} at 1.1 {\microns}, while otherwise the lowest detectable \ce{H2O} abundance is 1$\times10^{-4}$ VMR at all other \ce{CH4} values. When moving to longer wavelengths in Figure~\ref{fig:lineh2odetec}b, the low abundances of \ce{H2O} are only accessible at the lowest \ce{CH4} abundances, due to the dual molecular features breaking the spectral degeneracy.
 
Additionally, although shorter wavelengths are clearly preferred for dual \ce{CH4} and \ce{H2O} detection, the 0.9--1.1 {\microns} regime is burdened with increased coronagraph detector noise as detailed below, leading to longer required exposure times and thus higher required SNRs for detection.

\section{Conclusions \& Future Work}
\label{sec:conc}
\vspace{0.2cm}

\noindent
\fcolorbox{frame}{background}{%
    \parbox{\dimexpr\linewidth-2\fboxsep-2\fboxrule}{%
        \textbf{To Summarize:}
        \begin{gitemize}
            \item Modern levels of \ce{CH4} (1.65$\times10^{-6}$ VMR) are not detectable at any SNR$\le20$ with any bandpass width. Archean levels of \ce{CH4} (7.07$\times10^{-3}$ VMR) are easily detectable at all SNRs and bandpass widths.
            \item \ce{CH4} detectability is a function of the abundances of both \ce{CH4} and \ce{H2O}, due to the close overlap of spectral features for both species. This overlap between \ce{CH4} and \ce{H2O} absorption features becomes most acute at low to moderate \ce{CH4} abundances, and as the \ce{H2O} abundance increases, the required SNR for \ce{CH4} detection increases. 
            \item \ce{H2O} detectability depends on the abundance of \ce{CH4}; as \ce{CH4} abundance increases, the required SNR to detect low to moderate \ce{H2O} increases.
 
        \end{gitemize}
    }%
}
\vspace{0.2cm}

\ce{CH4} and \ce{H2O} detectability clearly relies heavily on the bandpass selected for observation. At longer wavelengths, detectability drops for both molecules, and instrument efficiency decreases as a result of moving further into the NIR. Shorter wavelengths remain preferable for detection, namely at 1.1 {\microns} for both \ce{CH4} and \ce{H2O}, and 0.9 {\microns} for \ce{H2O}. However, when searching for moderate to low \ce{CH4} in an \ce{H2O}-rich atmosphere, or searching for moderate to low \ce{H2O}in a \ce{CH4}-rich atmosphere, a higher SNR will be required to disentangle the molecular absorption signals. This may impact the science goals for HWO depending on whether an Archean Earth atmosphere or Modern Earth atmosphere drives the SNR requirement. 

In future works, the BARBIE project will culminate with an investigation of molecular detection across the full expected wavelength range for the Habitable Worlds Observatory. We will explore the requirements for molecular detection across the UV, optical, and NIR to determine the optimal wavelength range and SNR per molecule for strong detection. These metrics will be coupled with also varying the bandpass width across 20\%, 30\%, and 40\% bandpasses, with the new addition of a 10\% bandpass width. By using the KEN grids, we will explore molecular relationships through wavelength to investigate the most efficient observing strategy to detect multiple molecules with the most optimal bandpass wavelength, width, and SNR.

We also note that many of the critical wavelength bandpasses for optimal observation of \ce{H2O} fall at the intersection of the high-sensitivity regions of different detector technologies. This wavelength dependency could significantly impact the requirements for biosignature detection and result in significant differences to characterization yields, as shown by \citet{stark24b}. In future works, we will investigate how coronagraph detectors and their wavelength-dependent sensitivity can vary the detectability of different molecular species. 
\\

N. L. gratefully acknowledges financial support from an NSF GRFP and NASA FINESST. N.L. gratefully acknowledges Dr. Joseph Weingartner for his support and editing. N. L. also gratefully acknowledges Greta Gerwig, Margot Robbie, Ryan Gosling, Emma Mackey, and Mattel Inc.{\texttrademark} for Barbie (doll, movie, and concept), for which this project is named after. This Barbie is an astrophysicist! The authors would like to thank the Sellers Exoplanet Environments Collaboration (SEEC) and ExoSpec teams at NASA's Goddard Space Flight Center for their consistent support. MDH was supported in part by an appointment to the NASA Postdoctoral Program at the NASA Goddard Space Flight Center, administered by Oak Ridge Associated Universities under contract with NASA.

\bibliographystyle{aasjournal}
\bibliography{main}

\end{document}